%% file: geometry-vortex-v25.tex
\documentclass[12pt]{elsart}

\input{latex-macros.tex}

\newcommand{\arcl}{s}
\newcommand{\flex}{\mathrm{flex}} 
\newcommand{\cen}{\mathrm{c}} 
\newcommand{\oo}{o}
\newcommand{\ii}{\iota}
 
\newcommand{\details}[1]{}

\begin{document}

\begin{frontmatter}




\title{Vortex filament dynamics and \\ vortex ring motion revisited}


\author{Andrew D. Gilbert}

\address{Department of Mathematics and Statistics,\\
University of Exeter, Exeter, EX4 4QF, U.K.}

\begin{abstract}
\noindent
The motion of a slender vortex in ideal incompressible fluid is a classic problem in hydrodynamics. Formulae for the velocity of a vortex ring go back to work of Kelvin, Helmholtz, Hicks and Dyson in the nineteenth century, while more recently a number of models have been created for simulating the motion of slender tubes of vorticity of general shape. For a vortex tube or filament localised near to a curve $\Cc(t)$, the  relevant small parameter to measure slenderness is the tube radius divided by the radius of curvature of $\Cc$. The present paper revisits this range of classic problems by introducing a coordinate system closely linked to the geometry of vortex surfaces in a slender vortex. The motion of fluid elements in this coordinate system has an action--angle form, and with this the vorticity equation simplifies radically. At the same time, however, most aspects of the shape or evolution of a vortex are thrown into the description of the coordinates, and in particular the corresponding metric and volume form. 

As the coordinate system is non-orthogonal and time-dependent, tools of differential geometry are most easily used to describe the structure of both vorticity and coordinate system, and a general mathematical framework is set out. 
This resulting system of equations is taken as far as possible with only the assumption of vortex slenderness  in place, but allowing arbitrary motions and distortions of the vortex core and of the curve $\Cc(t)$. The modelling is applied to calculate the motion of a slender vortex ring with arbitrary axial flow, solving directly for the shape of perturbed vorticity surfaces and giving results in agreement with earlier studies. The general framework set up in this paper is suitable for the development of simplified equations for vortex motion and interaction in future studies. 

\end{abstract}

\end{frontmatter}



\section{Introduction}\label{secintro}

\noindent
Problems of fluid dynamics are often most naturally described in terms of vorticity $\omegav(\rv, t)$ rather than the fluid velocity $\uv(\rv, t)$. In the framework of ideal, incompressible flow, taking the curl of the Euler equation eliminates pressure and gives the vorticity equation, with a straighforward interpretation, thanks to Helmholtz, of how vorticity is transported in the flow. The long-range nature of the pressure field is, however, hidden in the elliptic problem of inverting vorticity to give the flow field, using the Biot--Savart law. Nonetheless this is an attractive framework, particularly when the vorticity is concentrated in a tube or filament, localised near a curve $\Cc(t)$, or several such tubes, to be transported and stetched by their induced flow.

The present paper is one of a series that applies geometric tools to problems of describing the motion of such slender vortices in ideal, incompressible fluid flow. The first paper \citep{ChGi21} described area waves \citep{LuAs89, Le94, Ma02} on a columnar vortex. These can be thought of as nonlinear, axisymmetric Kelvin waves where the cross-sectional area undergoes significant changes. Our goal in this paper is to set up a mathematical framework for the general evolution and motion of a slender vortex, and future papers will apply this framework, and also simplify it, for specific problems of vortex evolution and interaction. We will summarise some background and a few key papers; for many further references and discussion, see the books by \cite{Sa92},  \cite{AlKuOk07}, \cite{TiKlKn07} and \cite{Ma25}.

The classic mathematical problem in the field is to calculate the self-induced motion of a vortex ring with major radius $R$ (of curvature of $\Cc$) and minor radius $a$ (of the vortex tube) . Work of Kelvin, Helmholtz, Dyson and Hicks in the nineteenth century gave the propagation velocity involving a logarithmic term $\log R/a$; see \cite{Wi75}, \cite{Sa92} and \cite{ShLe92} for reviews and original references. These calculations are for vortex rings with axial vorticity, that is vorticity directed along the circle $\Cc$, but not axial flow along $\Cc$. Incorporating axial flow leads to a more complicated problem addressed by  \cite{Sa70} \citep[discussed in][]{Sa92} and \cite{WiBlZa71}, and results in the somewhat counter-intuitive finding that its presence acts to reduce the vortex ring velocity, and the propagation direction can even reverse if the axial flow is sufficiently strong. Other developments in the field of vortex rings include higher order calculations \citep{Fr70, Fr72}, structure of fat vortex rings \citep{No73}, spreading due to viscosity \citep{TuTi67,FuMo00,Fu02}, motion of helical vortex tubes \citep{Ri94,FuOk05}, vortex leap-frogging and interactions \citep{Ri98}, and waves and instabilities \citep{WiBl71,WiBlTs74,KoCh91,KoCh95}. 

The natural generalisation from a slender vortex ring is to a slender tube of vorticity localised near to a general curve $\Cc(t)$, and to ask how the curve, and the description of the vorticity inside the tube, evolve in time. There is a substantial literature on this problem, and two complementary approaches. The first is to think of a vortex tube as an object containing a fluid in motion, which responds to forces on its outer surfaces and which undergoes motion and stretching according to the balance of these forces. In this approach the local cross section of the tube is circular and profiles of the vorticity field (axial and azimuthal components) within the tube are typically assumed, for example uniform axial vorticity, uniform axial flow, and can undergo only limited changes in response to motion and stretching. Papers adopting this approach for modelling vortex motion include \cite{MoSa72} and \cite{Ma91,Ma92,Ma93}, and have the benefit of suppressing complexity in the core dynamics at the outset. 

The second approach, which is more closely aligned to the present study, is to make no assumptions about core structure but to set about a systematic expansion of the equations of fluid motion about the evolving curve $\Cc(t)$, the aim being to gain equations for both the internal vorticity profiles and for the motion of the curve. This approach was pioneered in a series of papers by Ting and collaborators \citep{TuTi67,Ti71,CaTi78,KlTi92}, with a comprehensive review in the book \cite{TiKlKn07}. The vorticity is assumed to be localised around a curve $\Cc(t)$ in space, and a local, orthogonal coordinate system is taken about the curve; the vortex is assumed to be slender so that the tube radius $a$ times the curvature $\kappa$ is small, $a \kappa = \eps \ll 1$. Weak viscosity may be included or not, but the focus of the present paper and this introduction is on the inviscid case. At leading order, in these coordinates, the flow is axisymmetric; however naturally the curvature of $\Cc$ brings in corrections of order $\eps$ and these can be treated using standard methods of matched asymptotic expansions, the vorticity-bearing flow within the tube being matched to the irrotational flow outside. Further developments include higher order expansions \citep{FuMi91}, PDEs for the self-stretching of filaments with short wavelength distortions \citep{KlMa91a,KlMa91b}, numerical simulations \citep{KlKn95,KlKnTi96}, stratified vortices \citep{RoMaKl24}, and waves on vortex filaments \citep{MaBr01,Ma02}. 

Given the wealth of studies in this field, what does the present paper offer? In order to give an answer, we should stress that much is unknown about vortices when they interact strongly: for example the numerical simulation of \cite{BuKe08} shows two vortex filaments forming a dipole of a hairpin shape moving under their mutual interaction. This leads to strong stretching, significant deformation of the vortex cores, the shedding of vorticity, and the possibility of blow-up of vorticity in a finite time 
\citep[see also, e.g.,][]{PuSi87,Pe01,BrHoPu16,MoKi19a,MoKi19b}. Such vortex dipoles have been studied in axisymmetric geometry \citep{Ri98,ChGiVa16}, where the stretching is algebraic in time, but less successfully in the general three-dimensional case where the outcome is still unclear \citep{ChGi18,ChGi25}. Other areas where analytical modelling is difficult include vortex reconnection \citep[e.g.][]{Sa90}, complex core dynamics \citep{MeHu94} and vortex breakdown \citep{Le78}. 

Our goal in this paper is to set up models for vortex filaments or tubes that allow very general dynamics and strong distortions of the core, with only the constraint being that a vortex must remain slender at all times. We adopt the machinery of differential geometry as set out in, for example, the books \cite{HaEl73}, \cite{Sc80}, \cite{Fr97} and \cite{ArKh98} \citep[see also][]{BeFr17}, and as recently used in revealing the geometric underpinning of generalised Lagrangian mean (GLM) theory of \cite{AnMc78a,AnMc78b}, by \cite{GiVa18,GiVa25}. The aim is to make the coordinate system absorb as much of the geometry of the vortex as possible, in particular vortex surfaces are given by a surfaces of constant coordinate $r$. This has the effect of simplifying the vorticity equation radically, but then throwing the complexity onto the description of the geometry. 

The equations we gain at the end of the paper (summarised in appendix \ref{appgensys}) are still quite general as the only assumption made in the derivation is that of a slender vortex.
To apply them to specific and novel configurations, we need to undertake further modelling and approximation relevant to a particular vortex geometry and dynamical process, and this will be the subject of future studies. However, as a test and to gain useful intuition, we apply the method to the velocity of a vortex ring with axial flow, regaining known results \citep{Sa70,Sa92,WiBlZa71}, but with the novelty that we solve directly for the shape of vorticity surfaces. In this application, our study has points of contact with \cite{Fr70,Fr72} for vortex ring motion without axial flow, and \cite{KoCh91,KoCh95} who study the stability of vortex rings to equivortical perturbations using a coordinate system similar to ours. Our coordinate system are also close related to the area coordinates used in \cite{Ch87} and \cite{ChGi25}, after contour averaging equations for axial velocity and vorticity.

The paper is structured as follows. In section 2 we set out intrinsic equations for the evolving shape of a curve $\Cc(t)$ in terms of the velocity $\Uv_\cen$ of points on the curve, together with two coordinate systems attached to the curve. We  write out the governing equations of vortex dynamics in a purely geometric form, and in components. We then consider fluid dynamics in the so-called \emph{vortex coordinate system} in \S3, which is that used by many authors, in particular we determin the metric and properties of the background flow, that is the velocity $\Uv$ of the coordinate system itself. Section 4 concerns the introduction of what we (inaccurately) term the \emph{Lagrangian coordinate system}: here the surfaces on which vortex lines spiral are surfaces of constant coordinate $r$, together with an angle coordinate $\theta$ and axial coordinate $z$. The exact equations for fluid dynamics in the Lagrangian coordinate system are unwieldy and so in \S5 we introduce the slender vortex limit and discuss simplifications to the metric and background flow properties. At this point we have a lot of things to deal with, and so to gain intuition we simplify to the problem of vortex ring motion, including axial flow, in \S6. We see how the equations governing vorticity and momentum lead to two differential equations whose solutions give directly the geometry of the vorticity distribution in the curved slender vortex, and lead to determination of its velocity from matching with a standard Biot--Savart calculation. In \S7 we continue with the general development without assuming slenderness; in \S8 we apply the slender vortex limit to gain systems of equations which are the main results of this paper. Finally section 9 offers concluding discussion, including future plans for applying the framework for problems of vortex motion and interaction.


\section{Governing equations and coordinate systems}\label{secgoveq}

\noindent
Our formulation is based on a vortex filament being localised close to a time-dependent curve $\Cc(t)$ lying in $\Rbb^3$. In this section we set out the various coordinate systems we need. We then discuss the geometric formulation of fluid dynamics that we will use, following the notation of \cite{ChGi21}.

\details{
Notation:
\begin{itemize}
\item
$\Uv$ is the background flow or velocity of the coordinate system
\item
$\uv$ is the full flow field 
\item
$\vv$ is the relative fluid flow 
\item
Lagrangian coordinates/space $(r, \theta ,z)$ or $(x,y,z)$, strictly Hybrid Euler--Lagrangian coordinates...
\item
Vortex coordinates/space $(\rt, \thetat,\zt)$ or $(\xt, \yt, \zt)$. 
\item
Ambient coordinates/space $(X,Y,Z)$ or $(R, \Theta, Z)$.
\end{itemize}
} 


\subsection{Curve dynamics}\label{sseccurve}

\begin{figure}
\centering
\includegraphics[scale=0.5]{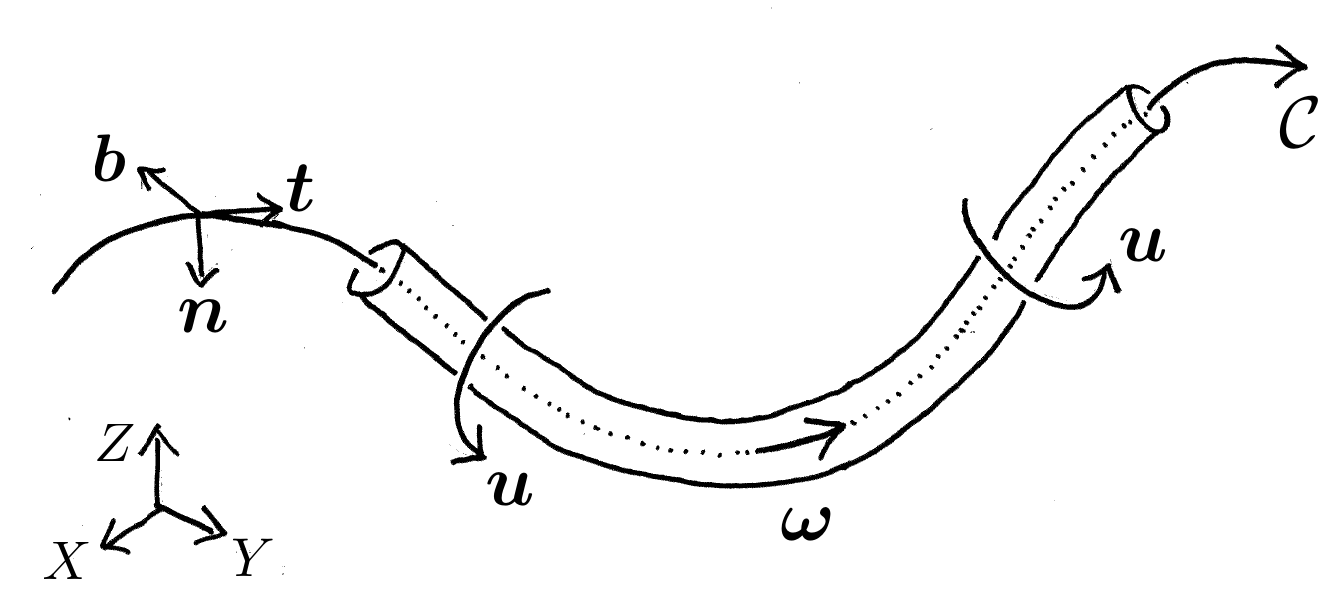}
\caption{Evolving curve $\Cc(t)$ in the ambient space $\Rbb^3$ with the Frenet--Serret basis $\{ \nv, \bv, \tv \}$. A slender vortex is localised near to $\Cc$.}
\label{figamb}
\end{figure}

\noindent
The vortex is localised near to a curve $\Cc(t)$ as depicted in figure \ref{figamb}. We will take $\Cc$ to be periodic to avoid consideration of any end conditions. We use $z$ as a Lagrangian coordinate along $\Cc$ and take points on the curve to be given by $\rv_\cen(z, t)$ in $\Rbb^3$. Cartesian coordinates in $\Rbb^3$ will be denoted $(X,Y,Z)$. 

Holding time constant, we have the Frenet--Serret equations 
\begin{equation}
\partial_{z} \rv_\cen   = J \tv, \quad 
\partial_{z} \tv = J \kappa \nv, \quad
\partial_{z} \nv = - J\kappa \tv + J\tau \bv, \quad
\partial_{z} \bv  = - J\tau \nv. 
\label{eqserretfrenet1}
\end{equation}
Here $J (z, t) = |\partial_{z}\rv_\cen|  = \partial s / \partial z$ is the Jacobian, where $\arcl$ is arclength, $\kappa(z,t)$ is the curvature and $\tau(z,t)$ is the torsion. We take the point on the curve at fixed $z$ to have velocity 
\begin{equation}
\partial_{t} \rv_\cen   |_{z}   = U_\cen\nv + V_\cen \bv + W_\cen \tv  = \Uv_\cen, 
\label{eqcurvevelocity1}
\end{equation}
with components $(U_\cen,V_\cen,W_\cen)(z,t)$ referred to the Frenet--Serret basis. We use the Roman font subscript `$\cen$' to denote quantities defined on the curve $\Cc$, which will later be extended off the curve; for example $\rv_\cen(z,t)$ is position on the curve whereas $\rv$ denotes a general position in space. In the same way, the velocity $\Uv_\cen$ or $(U_\cen,V_\cen,W_\cen)$ of points on the curve $\Cc$ will be extended to a flow $\Uv$ or $(U,V,W)$ in the neighbourhood of $\Cc$, something that can be done in many ways. The quantities $J$, $\kappa$, $\tau$ are not defined off the curve and so do not need the subscript `$\cen$', and likewise for the Frenet--Serret basis. Note that we have not yet introduced any information about a  fluid flow, and so the velocity  $\Uv_\cen$ is arbitrary at this point. 

The curve velocity in (\ref{eqcurvevelocity1}) also gives the time evolution of the basis $\{ \nv, \bv, \tv\}$ and the intrinsic geometry represented in $\{J, \kappa, \tau\}$. A standard calculation yields
\begin{subequations}\label{eqcurvebasisdt1}
\begin{align}
\partial_{t} J |_{z} & = J(W_{\cen\arcl} - \kappa U_\cen) ,  \\
\partial_{t} \kappa |_{z} & = (U_{\cen\arcl} - \tau V_\cen)_\arcl  - \tau(V_{\cen\arcl} + \tau U_\cen)    + \kappa_\arcl W_\cen + U_\cen \kappa^2 , \\
\partial_{t} \tau  |_{z} & = S_{\cen\arcl}  + \kappa (V_{\cen\arcl} + \tau U_\cen) - \tau(W_{\cen\arcl} - \kappa U_{\cen}), 
\end{align}
\end{subequations}
with  
\begin{subequations}\label{eqcurvebasisdt2}
\begin{align}
\partial_{t} \nv|_{z}  & = S_\cen \bv - ( U_{\cen\arcl} - \tau V_\cen  + \kappa  W_\cen) \tv,  \\
\partial_{t} \bv |_{z} & = - S_\cen\nv -  (V_{\cen\arcl} + \tau U_\cen) \tv , \\
\partial_{t} \tv |_{z} & = (U_{\cen\arcl}- \tau V_\cen  + \kappa W_\cen) \nv + (V_{\cen\arcl} + \tau U_\cen) \bv .
\end{align}
\end{subequations}
We have introduced an expression that appears often in our calculations without simplifying appreciably, 
\begin{equation}
S_\cen  = \kappa^{-1} (V_{\cen\arcl} + \tau U_\cen)_\arcl + \kappa^{-1}\tau ( U_{\cen\arcl} - \tau V_\cen ) + \tau W_\cen. 
\label{eqS0def} 
\end{equation}
We sometimes refer to $S_\cen(z,t) $ as the \emph{spin}, namely the angular veocity of the $\{\nv, \bv\}$ axes about the tangent vector $\tv$  at a fixed $z$ location on the curve $\Cc$.

\details{
Also useful can be the intermediate components of the calculations: 
%
\begin{align*}
\partial_{t} (J \kappa) |_{z} & = J(U_{\cen\arcl} - \tau V_\cen + \kappa W_\cen)_\arcl - J\tau(V_{\cen\arcl} + \tau U_\cen) , \\
\partial_{t} (J\tau)  |_{z} & = JS_{\cen\arcl}  + J \kappa (V_{\cen\arcl} + \tau U_\cen). 
 \end{align*}
%
}

As seen above, to denote derivatives we use both letter subscripts on the partial derivative symbol, for example $\partial_t$ in (\ref{eqcurvebasisdt1}, \ref{eqcurvebasisdt2}), and also subscripts such as $s$ in $\kappa_\arcl$ or $U_{\cen\arcl}$, according to emphasis and visual appeal. We stress that the time derivative  $\partial_{t}$ on the left-hand side of (\ref{eqcurvebasisdt1}, \ref{eqcurvebasisdt2}) is taken at fixed Lagrangian coordinate $z$, even though for compactness, on the right-hand side it is convenient to write derivatives along the curve with respect to arc length $\arcl$ rather than~$z$ itself. For any time-dependent function $F(z,t)$ defined on $\Cc$, derivatives are related by the chain rule
\begin{equation}
\partial_{z} F(z,t)  = J \, \partial_\arcl F(z,t) , 
\label{eqszswap} 
\end{equation}
so we may freely replace  $\partial_\arcl$ by $J^{-1} \partial_z$ and \emph{vice versa} in the above differential equations.

Plainly moving points along the curve with velocity $W_\cen$ does not change the actual curve, only its Lagrangian labelling, and this gives the gauge freedom to relabel. If we introduce a new coordinate $\zh$, then equations (\ref{eqcurvebasisdt1}) are invariant under the replacement of $(z,t)$ with $(\zh,\tth)$, $J$ with $\Jh$, and $W_\cen$ with $\Wh_\cen$, where
\begin{equation}
\zh = \zh(z,t), \quad \tth = t , \quad \Jh = \frac{\partial s}{\partial \zh} \Big|_{\tth}\, , \quad 
\Wh_\cen = W_\cen - \Jh \, \frac{\partial \zh}{\partial t} \Big|_{z}\, .
\label{eqgauge1} 
\end{equation}
In the next subsections we introduce the three coordinate systems that we will use, building up in terms of complexity.

\subsection{Ambient coordinate system}\label{ssecambcoord}

\noindent
In the Euclidean space $\Rbb^3$ containing the fluid flow, we will use either Cartesian coordinates $(X,Y,Z)$ or cylindrical polar coordinates $(R, \Theta, Z)$ defined in the usual way, and refer to either of these as the \emph{ambient coordinate system}. 
In this \emph{ambient space} we have the usual metric and volume form 
\begin{subequations}\label{eqgmuambient1}
\begin{align}
\gv  &  = dX^2 + dY^2 + dZ^2 = dR^2 + R^2\, d\Theta^2 + dZ^2, \\
\muv  & = dX\wedge dY \wedge dZ = R \, dR\wedge d\Theta \wedge dZ, 
\end{align}
\end{subequations}
and with unit vectors $\{ \iv, \jv , \kv\}$ the position vector is
\begin{equation}
\rv = X  \iv + Y  \jv + Z \kv = R \cos \Theta\, \iv + R \sin \Theta\, \jv + Z \kv. 
\end{equation}
We also define a time variable $T \equiv t$ so that $\partial_T$ means derivative at constant $(X,Y,Z)$. 


\subsection{Vortex coordinate system}\label{sseccortexcoord}

\noindent
Our second coordinate system moves in the ambient space following the vortex centre line $\Cc$. It makes use of the Frenet--Serret basis to define a coordinate plane perpendicular to the centreline curve $\Cc$ at each Lagrangian station labelled by $z$. This coordinate system has been used in many studies since \cite{Ti71} and \cite{MoSa72}, and is a key stepping stone to our third and final coordinate system.

We let $\zt \equiv z$ denote the Lagrangian coordinate on $\Cc$ as in \S2.1 and write the coordinate system  as $(\xt, \yt, \zt)$ or as $(\rt, \thetat, \zt)$. We use the Serret--Frenet basis $\{\nv, \bv, \tv\}$ along the curve $\Cc$ as in (\ref{eqserretfrenet1}) and set 
\begin{align}
\rv & = \rv_\cen(\zt, \ttt) + \xt\,  \nv (\zt, \ttt) + \yt \, \bv(\zt, \ttt).
\label{eqrvvortecoord1}
\end{align}
We can also set 
\begin{align}
\xt =  \rt \cos \thetat , \quad \yt = \rt \sin \thetat .
\label{eqxtytrtthetat}
\end{align}
We refer to either of these coordinate systems, $(\xt, \yt, \zt)$ or $(\rt, \thetat, \zt)$, as the \emph{vortex coordinate system}. For simplicity, but with some abuse of terminology, we will refer to $(\xt,\yt,\zt)$ as the \emph{Cartesian} form of the coordinates, which we will use primarily, while $(\rt, \thetat, \zt)$ give the corresponding \emph{polar} form. On the curve $\Cc$ itself we have $\xt = \yt = \rt = 0$. We also introduce a time variable $\ttt \equiv t \equiv T $ so that $\partial_{\ttt}$ denotes a time derivative at fixed $(\xt, \yt, \zt)$ coordinates.  This coordinate system and the next one inherit the metric $\gv$ and volume form $\muv$ from those in the ambient space in (\ref{eqgmuambient1}). We will write these down later in the paper. 

Note that as far as points on $\Cc$ are concerned the coordinate pairs $(\zt, \ttt)$ and $(z, t)$ are identical, and so equations 
(\ref{eqserretfrenet1}--\ref{eqgauge1}) for curve geometry and evolution are the same under the interchange $(z,t)\leftrightarrow(\zt,\ttt)$. There is no need to allow a more complicated link between $z$ and $\zt$, as this is already included in the gauge freedom set out  in (\ref{eqgauge1}). 


\subsection{Lagrangian coordinate system}\label{sseclagcoord}

\noindent
Our third coordinate system also moves with the vortex in the ambient space but will be more closely adapted to the surfaces defined by vortex lines, as we will explain later. We refer to this for brevity as the \emph{Lagrangian coordinate} system, though it is technically  described as \emph{hybrid Eulerian--Lagrangian}, as in for example \cite{SoRo10}. Again we use a Cartesian version as $(x,y,z)$ or polar as $(r, \theta, z)$, related by 
\begin{align}
x =  r \cos \theta , \quad y = r \sin \theta , 
\end{align}
with a corresponding time variable $t$ and $\partial_t$ denoting a derivative at fixed location in this system. 
We again peg the coordinate system to the centre line curve $\Cc$; we set $x=y=r = 0$ and $ z \equiv \zt$ on $\Cc$ as before, so that the curve evolution equations (\ref{eqserretfrenet1}--\ref{eqgauge1}) continue to hold.


As we move off $\Cc$ we allow the new coordinate system to flex in ways that will follow the vortex geometry. We will make most use of the polar version $(r,\theta,z)$ and we set  
\begin{align}
\rv   
 = \rv_\cen(\zt(r,\theta,z,t), t) 
& + \xt(r,\theta,z,t)\,  \nv (\zt(r,\theta,z,t), t)  + \yt(r,\theta,z,t) \, \bv(\zt(r,\theta,z,t),t) .
\label{eqrvlagcoord1cart}
\end{align}
Here the vortex coordinates $(\xt,\yt, \zt)$ are now taken as arbitrary time-dependent functions of the Lagrangian coordinates $(r,\theta, z)$ subject only to their behaviour on $\Cc$, 
\begin{equation}
\xt(0,\theta,z,t) = \yt(0,\theta,z,t) = \rt(0, \theta,z,t) =0, \quad \zt(0,\theta,z,t) = z. 
\label{eqoriginset}
\end{equation}
These functions will be defined so that the coordinate system follows the geometry of vorticity surfaces, as discussed in \S\ref{seclagcoord}.


\subsection{Geometric formulation of fluid dynamics}\label{ssecgeometricfull}

\noindent
We will now set out a general geometric formulation of vortex dynamics, following the notation in \cite{ChGi21}. This can be applied to any coordinate system, allowed to be non-orthogonal and moving in the ambient space, for example in any of those introduced above, that is $(X,Y,Z,T)$, $(R, \Theta, Z, T)$, $(\xt, \yt, \zt, \ttt)$, $(\rt, \thetat, \zt, \ttt)$, $(x,y,z,t)$ or $(r,\theta, z, t)$. 

For the case of a moving coordinate system (the last four of these six) we have three velocity fields to deal with. The first is the velocity of a fluid element in the ambient space, which is the `true' \emph{fluid velocity} and we denote this by $\uv$. However when we use a moving coordinate system, a point with fixed coordinates in this system will move in the ambient space with a \emph{background velocity} which we write as $\Uv$. Finally, with respect to the moving coordinate system a fluid element has a \emph{relative velocity} of $\vv$. These three velocity fields are then simply related by  $\uv = \Uv + \vv$. For a steady coordinate system (the first two of the above six) we simply set $\Uv = 0$, $\uv = \vv$ and regain Eulerian fluid mechanics. If we instead set $\Uv = \uv$, $\vv=0$ we have purely Lagrangian fluid mechanics; the coordinate system follows fluid particles. Otherwise we have the possibility of choosing an arbitrary division between these two, by allowing $\Uv$ to take up part of the fluid motion, and $\vv$ to take up the other part, hence the term \emph{hybrid Eulerian--Lagrangian}. As in many earlier studies, the background flow $\Uv$ can take up the motion of the filament centre line $\Cc$ and its neighbourhood, and the flow $\vv$ capture the spinning of fluid elements about $\Cc$.

The general equations for fluid dynamics in a vorticity formulation may be written as 
\begin{subequations}\label{eqgengeom}
\begin{align}
 & \partial_t \xiv + d (\vv \ip \xiv) = 0 ,  \label{eqgengeom10}\\
 & \xiv = d \nuv,  \label{eqgengeom20}\\
 & \nuv = \uv_\flat \equiv  \gv(\uv, \cdot) ,  \label{eqgengeom30}\\
  & d(\uv \ip \muv) \equiv \muv \divv \uv = 0 ,  \label{eqgengeom40}\\ 
 & \uv = \Uv + \vv ,    \label{eqgengeom50}
\end{align}
\end{subequations}
and we discuss these in turn. 
Equation (\ref{eqgengeom10}) is the vorticity equation written in the language of differential geometry, giving transport of the vorticity two-form field $\xiv$ in the relative flow field $\vv$. Note that the key term $d (\vv \ip \xiv)$ can also be written as the Lie derivative $\mathcal{L}_{\vv} \xiv$ (using Cartan's formula), indicating its nature as a transport term. 
Equation (\ref{eqgengeom20}) relates vorticity $\xiv$ to the fluid momentum one-form field $\nuv$ using the exterior derivative $d$. The geometry of the coordinate system comes in through the presence of the metric $\gv$ and volume form $\muv$: equation (\ref{eqgengeom30}) relates momentum $\nuv$ to the fluid velocity vector field $\uv$ using the metric, while equation (\ref{eqgengeom40}) expresses that the fluid velocity $\uv$ is incompressible. Finally, while the vorticity $\xiv$ is related to the incompressible fluid velocity $\uv$ through (\ref{eqgengeom}b--d), we need to subtract the background velocity $\Uv$ from $\uv$ to obtain the relative velocity $\vv$ that actually transports vorticity in (\ref{eqgengeom10}), and this is expressed in (\ref{eqgengeom50}). 

We stress that the system (\ref{eqgengeom})  applies in any coordinate system, through its geometry being encoded in the metric $\gv$ in (\ref{eqgengeom30}) and volume form $\muv$ in (\ref{eqgengeom40}), while any time-dependence is expressed by $\Uv $ in (\ref{eqgengeom50}). From the point of view of the discussion and pictures, it is often convenient to think about the vorticity as a vector field $\omegav$, given by $\omegav\ip \muv = \xiv$, but for calculations we will find the two-form $\xiv$ most useful. 


\begin{figure}
\centering
\includegraphics[scale=0.4]{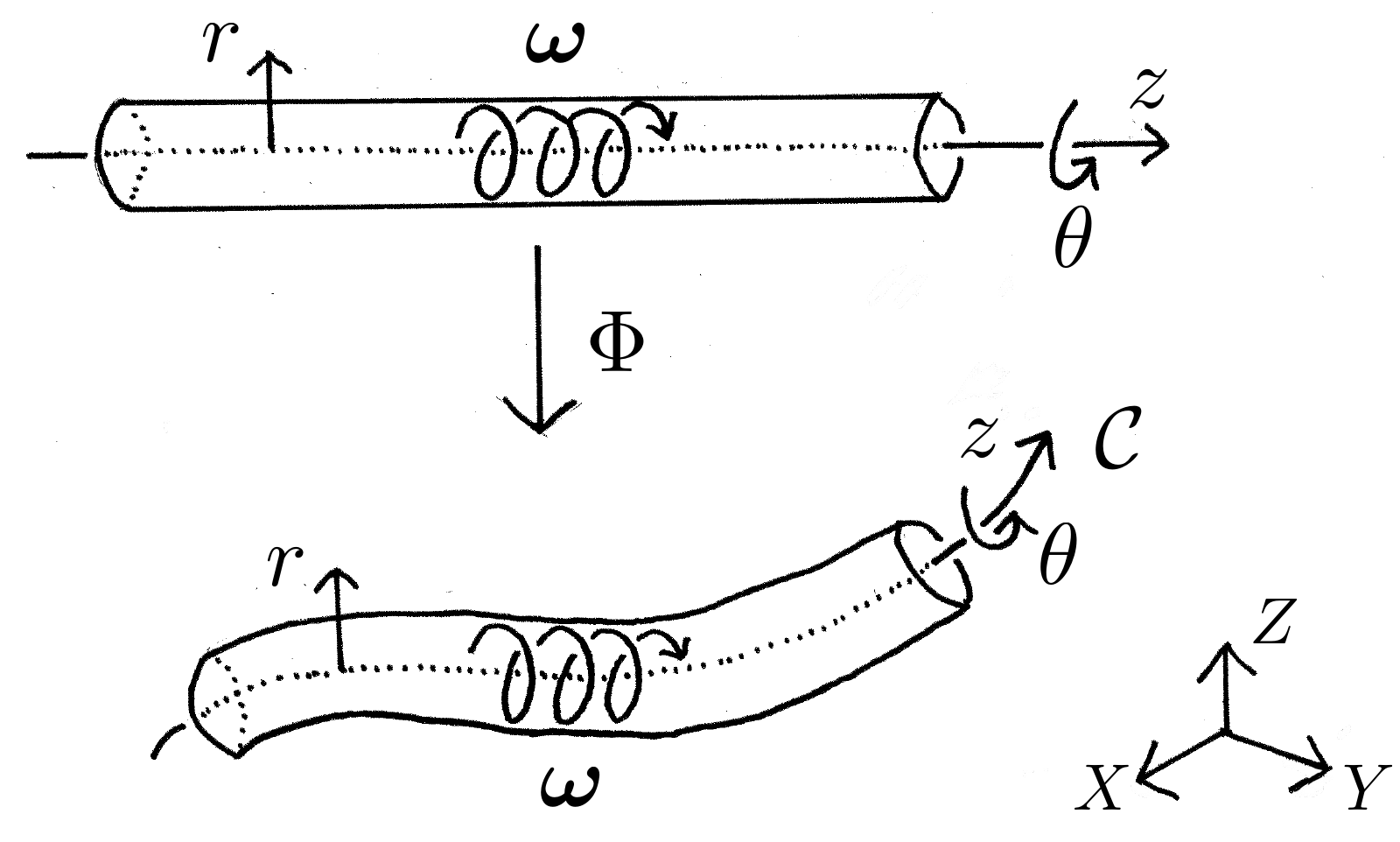}
 \caption{Mapping $\Phi$ from a vortex in the reference space with Lagrangian coordinates $(r,\theta, z,t)$ to the ambient space $(X,Y,Z,T)$.}
 \label{fig2}
\end{figure}

One way of viewing this system of equations, which can be helpful, is to look ahead to our use of the Lagrangian coordinate system $(r,\theta, z,t)$. We can think of a vortex as described in a \emph{reference space} with these as the usual cylindrical coordinates. This is shown at the top of figure \ref{fig2}, where vortex lines lie on cylinders of constant $r$ coordinate. The evolution of vorticity is relatively straightforward in the reference space as each vortex line remains wound on the same cylinder although amount of winding can vary in $z$ and in time $t$, in other words there can be non-uniform angular velocity. 

However fluid dynamics does not take place in the reference space, and so we suppose there is a time-dependent map $\Phi$ which then embeds the vortex in the ambient space, with coordinates $(X,Y,Z,T)$. The coordinates $(r,\theta, z, t)$ in the reference space are pushed forward by $\Phi$ to give moving, vortex-following coordinates in the ambient space as depicted. If we write down the equations of motion in the ambient space in a geometric form, and then apply a pull-back $\Phi^*$ to the reference space, we gain the system of equations (\ref{eqgengeom}) giving fluid dynamics in terms of $(r, \theta, z, t)$ \cite[see discussion in][]{ChGi21}. In reality, we need not have introduced a reference space, and we will scarcely mention it again, but simply think of $(r, \theta, z, t)$ as a moving coordinate system in the ambient space. Nonetheless the picture in figure~\ref{fig2} does show our approach in this study, in particular taking a simple structure of vorticity at the top, embedded into physical space at the bottom, and indicates the derivation of (\ref{eqgengeom}). 

One application of this point of view relates to the case of a purely Lagrangian coordinate system, so that $\Uv=\uv$ and $\vv=0$ meaning that $\partial_t \xiv =0$ from (\ref{eqgengeom10}) and vorticity is unchanging in the reference space. The vorticity in the ambient space at all times is then the initial vorticity at time $t=0$, pushed forward by the flow map $\Phi$ from time zero to time $t$, giving the Cauchy solution to the vorticity equation \cite[see][]{BeFr17}.

\subsection{Geometric formulation in components}\label{ssecgeometriccomp}

\noindent
We will need to write out the component version of the governing equations (\ref{eqgengeom}) and for ease of explanation we use coordinates $(r,\theta, z, t)$ in the rest of this section, Although these coordinates naturally have the flavour of cylindrical polar coordinates, we stress that the equations below in this subsection hold for any coordinate system. We represent vector fields such as $\uv$ by (contravariant) components, 
\begin{equation}
\uv = u^1 \partial_r + u^2 \partial_\theta + u^3\partial_z,  
\label{eqcontravariant}
\end{equation}
one-forms such as $\nuv$ by (covariant) components,
\begin{equation}
\nuv = \nu_1 \,dr + \nu_2  \,d\theta + \nu_3\, dz,  
\label{eqcovariant}
\end{equation}
and vorticity two forms such as $\xiv$ by
\begin{equation}
\xiv = \xi_1 \, d\theta \wedge dz + \xi_2 \, dz\wedge dr + \xi_3 \,dr \wedge d\theta. 
\label{eq2form}
\end{equation}
It is convenient to write the volume element as 
\begin{equation}
\muv = \mu \, dr \wedge d\theta \wedge dz,  
\label{eqmudef0}
\end{equation}
so defining the quantity $\mu$. We can also note that the vorticity vector field $\omegav$ corresponding to the vorticity two-form field $\xiv$ has components given by $\xiv = \omegav \ip \muv$ or (cf.\ (\ref{eqcontravariant})), 
\begin{equation}
\omega^1 = \mu^{-1} \xi_1 , \quad
\omega^2 = \mu^{-1} \xi_2 , \quad
\omega^3 = \mu^{-1} \xi_3 .
\label{eqxitoomega}
\end{equation}

In components the first of the governing equations, the vorticity equation (\ref{eqgengeom10}), is
\begin{subequations}\label{eqvortcomp1}
\begin{align}
& \bigl(\partial_{t} + v^1 \partial_{r} + v^2 \partial_{\theta} + v^3 \partial_{z}\bigr) \xi_1 - \bigl(\xi_1\partial_{r} + \xi_2 \partial_{\theta} + \xi_3 \partial_{z}\bigr) v^1 + \bigl(v^1_{r} + v^2_{\theta} + v^3_{z})\xi_1 = 0  , \\
& \bigl(\partial_{t} + v^1 \partial_{r} + v^2 \partial_{\theta} + v^3 \partial_{z}\bigr) \xi_2 - \bigl(\xi_1\partial_{r} + \xi_2 \partial_{\theta} + \xi_3 \partial_{z}\bigr) v^2 + \bigl(v^1_{r} + v^2_{\theta} + v^3_{z}\bigr)\xi_2 = 0 , \\ 
& \bigl(\partial_{t} + v^1 \partial_{r} + v^2 \partial_{\theta} + v^3 \partial_{z}\bigr) \xi_3 -  \bigl(\xi_1\partial_{r} + \xi_2 \partial_{\theta} + \xi_3 \partial_{z}\bigr) v^3 + \bigl(v^1_{r} + v^2_{\theta} + v^3_{z}\bigr)\xi_3 = 0  , 
\end{align}
\end{subequations}
and gives the evolution of vorticity $\xiv$ components  in the relative flow field $\vv$. This vorticity $\xiv$ is linked to the momentum $\nuv$  by (\ref{eqgengeom20}) which may be written out as 
\begin{subequations}\label{eqzetaagain}
\begin{align}
\xi_1 & = \partial_{\theta} \nu_{3} - \partial_{z}\nu_{2} , \\
\xi_2 & = \partial_{z}\nu_{1} - \partial_{r}\nu_{3} , \\
\xi_3 & = \partial_{r}\nu_{2} - \partial_{\theta}\nu_{1}  .
\end{align}
\end{subequations}
We note that as a consequence of its definition in (\ref{eqgengeom20}), $d\xiv=0$, and so these components are not independent, but are related by  
\begin{equation}
\partial_r \xi_1 + \partial_\theta \xi_2 + \partial_z \xi_3 = 0 . 
\label{eqdxizero}
\end{equation}
The momentum $\nuv$ is linked to the full flow $\uv$ by the metric $\gv$ of the appropriate coordinate system, using (\ref{eqgengeom30}),
\begin{equation}
\nu_i = g_{ij} u^j  \qquad (i = 1, 2, 3). 
\label{eqnuagain}
\end{equation}
Obtaining the flow $\uv$ from the vorticity $\xiv$ is an elliptic problem which must be solved with the volume conserving condition  (\ref{eqgengeom40}) applying which is, from (\ref{eqmudef0}),
\begin{equation}
\partial_{r} (\mu u^1) + \partial_{\theta} (\mu u^2) + \partial_{z} (\mu u^3)= 0 . 
\label{eqdivagain}
\end{equation}
The motion of the coordinate system in the ambient space is captured by the background flow $\Uv$. The full flow is given by the sum of the background flow $\Uv$ and the relative flow $\vv$ as per (\ref{eqgengeom50}), 
\begin{equation}
u^i = U^i + v^i   \qquad (i = 1, 2, 3). 
\label{equUvagain}
\end{equation}

In the ambient space, 
 the metric $\gv$ and volume form $\muv$ are independent of time with $\partial_T \gv =0$, $\partial_T \muv =0 $, where the $\partial_T$ derivative denotes differentiation at a fixed (ambient) point $(X,Y,Z)$. Taking now a fixed location, say $(r,\theta,z)$, with respect to the moving coordinate system, we obtain equations for the time-evolution $\partial_t \gv$ of the metric and $\partial_t \muv$ of the volume form in this moving frame, namely, 
\begin{align}
0 = \partial_T \gv  &  = \partial_t \gv - \lie_{\Uv} \gv  , \quad 
0 = \partial_T \muv   = \partial_t \muv - \lie_{\Uv} \muv = \partial_t \muv - (\divv \Uv) \muv .
\end{align}
Note that these will be satisfied automatically provided $\gv$, $\muv$ and $\Uv$ are calculated correctly, and so these are not conditions that need to be imposed. However making use of (\ref{eqmudef0}) gives the useful identity, 
\begin{equation}
\divv \Uv = \mu^{-1} \partial_t \mu. 
\label{eqmuidentity}
\end{equation}

While the geometric formulation (\ref{eqgengeom}), written out in (\ref{eqvortcomp1}--\ref{equUvagain}), is very general, clearly it can become complicated once a specific metric $\gv$, volume form $\muv$ and background flow $\Uv$ are substituted. On the other hand, in this set up, each quantity has a clear meaning, namely velocity, momentum, vorticity, and we do not use covariant derivatives, or equivalently Christoffel symbols, which makes the approach geometrically based and relatively intuitive.


\section{Fluid dynamics in the vortex coordinate system}\label{secvortcoord}

\noindent
We have defined and discussed three coordinate systems in \S2, one static and two moving, together with a fluid dynamical formulation relevant to any of them in \S3. The first moving coordinate system  is given by the vortex coordinates $(\xt, \yt, \zt, \ttt)$ or $(\rt, \thetat, \zt, \ttt)$ of \S\ref{sseccortexcoord} and has been used by many authors. In this section we write down the machinery, for example metric $\gv$ and volume form $\muv$, that we need to express vortex dynamics in these coordinates, using our approach. We will however stop short of writing down everything we might use, because in this system the coordinate surfaces, particular of constant $\rt$, only \emph{approximately} follow vorticity surfaces and we aim to follow them \emph{exactly}. Use of the vortex coordinate system in \S\ref{sseccortexcoord} is then a vital stepping stone before our analysis in \S4 in which we set out fluid dynamics in the second set of moving coordinates $(r, \theta, z, t)$ of \S\ref{sseclagcoord}, where vorticity surfaces are given \emph{precisely} by surfaces of constant $r$, as expressed in figure \ref{fig2}.



\subsection{General framework}\label{ssecvortcoordgen}

\noindent
We work in the vortex coordinate system in the Cartesian form $(\xt, \yt, \zt, \ttt)$ given  in (\ref{eqrvvortecoord1}). From (\ref{eqcurvevelocity1}) and (\ref{eqcurvebasisdt2}) we find that the velocity in the ambient space of any point with fixed vortex coordinates $(\xt, \yt, \zt)$ is 
\begin{align}
\Uv & = \partial_{\ttt}\, \rv =   U_\cen\nv + V_\cen\bv + W_\cen\tv + \xt\,  \partial_{\ttt} \, \nv  + \yt \,\partial_{\ttt} \, \bv  =  U \nv + V \bv + W \tv  , 
\label{eqfullU}
\end{align}
with components  
\begin{subequations}\label{eqfullUi}
\begin{align}
U &  = U_\cen  - S_\cen \yt,  
  \label{eqfullU0}\\
 V & = V_\cen + S_\cen \xt ,  
  \label{eqfullU1}\\
 W & = W_\cen h  - ( U_{\cen\arcl} - \tau V_\cen)  \xt  - (V_{\cen\arcl} + \tau U_\cen)\yt . 
 \label{eqfullU2}
\end{align}
\end{subequations}
Here we define the useful quantity $h(\xt, \zt, \ttt)$ given by 
\begin{equation}
h = 1 - \kappa \xt  , 
\label{eqhdef}
\end{equation}
while the spin $S_\cen$ was given in (\ref{eqS0def}). Recall that $\Uv$ is the background flow, that is the velocity of the $(\xt, \yt, \zt)$ coordinate system. The coordinate system flexes through the motion $\Uv\!_\cen$ of the centre line curve $\Cc$ in (\ref{eqcurvevelocity1}), and the choice of coordinates fixes the extension of $\Uv\!_\cen$ to give the full background flow field $\Uv$ off the curve $\Cc$. Note that we have yet to specify a relationship between the curve velocity $\Uv\!_{\cen}$ and the actual fluid velocity $\uv$. 

While we have written down  $\Uv$, it is given in terms of the $\{\nv, \bv , \tv\}$ basis in (\ref{eqfullU}), and to use the usual machinery of differential geometry we need to refer this to the basis $\{\partial_{\xt}, \partial_{\yt}, \partial_{\zt}\}$ (cf.\ (\ref{eqcontravariant})). In other words we need the contravariant components $(U^1, U^2, U^3)$, with 
\begin{equation}
\Uv = U \nv + V \bv + W\tv  = U^1 \partial_{\xt} + U^2 \partial_{\yt}  + U^3 \partial_{\zt}.
\label{eqUVnbt123}
\end{equation}
To avoid a proliferation of notation in this section, we will not annotate components $U^i$ with tildes in (\ref{eqUVnbt123}) and similarly for components of other quantities, though it would be logical to do so. The context should make it clear what is going on.  To obtain the appropriate component transformation matrix, from $(U,V,W)$ to $(U^1, U^2, U^3)$, we  need to differentiate $\rv$ in (\ref{eqrvvortecoord1}) with respect to small changes in the coordinates $(\xt,\yt, \zt)$, which will result in (\ref{eqdrvbasic}) below.

However, before we set out the details of this calculation of the $U^j$, it is useful to exhibit the general structural information we need. We express the change in $d\rv$ with respect to the changes $(d\xt, d\yt, d\zt)$ in the coordinates by means of a matrix $\Qv$ defined by 
\begin{equation}
d\rv = \begin{pmatrix} d\xt & d\yt & d\zt \end{pmatrix}  \Qv \begin{pmatrix} \nv \\ \bv \\ \tv \end{pmatrix}. 
\label{eqgendrq}
\end{equation}
%
From the matrix $\Qv$ we gain the components of the metric $\gv$ and the corresponding volume form $\muv$, with 
\begin{equation}
\gv =  \begin{pmatrix} d\xt & d\yt & d\zt \end{pmatrix} \Qv \Qv^T \begin{pmatrix} d\xt \\ d\yt \\ d\zt \end{pmatrix} ,  \quad \muv =\mu \, d\xt \wedge d\yt \wedge d\zt, \quad
\mu =  \det \Qv . 
\label{eqgengqq}
\end{equation}
That is, the components of the matrix $\Qv\Qv^T$ are the components $g_{ij}$ of the metric $\gv$, and the determinant of the metric is $\mu^2$. 

We can apply the metric to go from the velocity field $\uv$ to its momentum $\nuv = \uv_{\flat}$ in (\ref{eqgengeom30}) or (\ref{eqnuagain}) and from the momentum $\nuv$ we obtain the vorticity $\xiv$ by applying the exterior derivative $d$ in (\ref{eqgengeom20}) or (\ref{eqzetaagain}). We can apply similar operations to the background flow $\Uv$ and to the relative flow $\vv$, and define
\begin{subequations}\label{eqschema}
\begin{align}
\text{fluid velocity} \  \uv \xrightarrow{\quad\flat\quad} &\ \text{momentum}\   \nuv  \xrightarrow{\quad d\quad} \text{vorticity}  \ \xiv , 
\\
\text{background velocity} \  \Uv \xrightarrow{\quad\flat\quad} &\ \text{momentum}\   \Upsilonv  \xrightarrow{\quad d\quad} \text{vorticity}  \ \Xiv , 
\\
\text{relative velocity} \  \vv \xrightarrow{\quad\flat\quad} &\ \text{momentum}\   \piv  \xrightarrow{\quad d\quad} \text{vorticity}  \ \varpiv  .
\end{align}
\end{subequations}
%
It is also useful to set out explicitly the obvious relationships
\begin{equation}
\uv = \Uv + \vv, \quad 
\nuv = \Upsilonv + \piv = \Upsilonv + \vv_\flat, \quad
\xiv = \Xiv + \varpiv = \Xiv + d \vv_\flat. 
\label{eqschemarels}
\end{equation}
Focusing on the background velocity $\Uv$, say, with the notation established in (\ref{eqUVnbt123}), the matrix $\Qv$ then relates components of background flow and momentum according to
\begin{equation}
\begin{pmatrix}
U \\  V \\  W
\end{pmatrix}
= 
\Qv^T 
\begin{pmatrix}
U^1 \\ U^2 \\ U^3
\end{pmatrix}, 
\quad\quad
\begin{pmatrix}
U^1\\ U^2\\ U^3
\end{pmatrix}
= 
(\Qv^{-1})^T 
\begin{pmatrix}
U\\ V\\ W
\end{pmatrix}, \quad\quad
\begin{pmatrix}
\Upsilon_1 \\ \Upsilon_2 \\ \Upsilon_3 
\end{pmatrix}
= 
\Qv
\begin{pmatrix}
U\\ V\\ W
\end{pmatrix}, 
\label{eqcoordgenq}
\end{equation}
and likewise for the fluid and relative velocities. 
  

\subsection{Metric and background flow}\label{ssecvortcoordback}

\noindent
Returning to our  development in detail, we differentiate (\ref{eqrvvortecoord1}) with respect to variations in the coordinates  to give 
\begin{equation}
d\rv =    (d\xt - J \tau \yt\, d\zt)  \nv +  (d\yt  +  J \tau \xt\, d\zt)  \bv + J h  \, d\zt\, \tv, 
\label{eqdrvbasic}
\end{equation}
using the Frenet--Serret equations (\ref{eqserretfrenet1}) (with $(\zt,\ttt) = (z,t)$ on $\Cc$). This is written as (\ref{eqgendrq}) with 
\begin{equation}
\Qv =
\begin{pmatrix}
1 & 0 & 0 \\
0 & 1 & 0 \\
\;- J \tau \yt \; & \;J \tau \xt \; & \;Jh\;
\end{pmatrix}, 
\quad
(\Qv^{-1})^T =
\begin{pmatrix}
\;1 \;& \;0\; & h^{-1} \tau \yt \; \\
0 & 1 & \;- h^{-1} \tau \xt \;\\
0  & 0  & \;J^{-1} h^{-1}\;
\end{pmatrix} .
\label{eqQvdef} 
\end{equation}
Using (\ref{eqgengqq}), the matrix giving the metric $\gv$ in terms of $(d\xt, \d\yt, d\zt)$ has components 
\begin{equation}
(g_{ij}) = 
 \begin{pmatrix}
\;1\; & \;0\; & - J \tau \yt \\
0 & 1 & J \tau \xt \\
\;- J \tau \yt\;  &  \;J \tau \xt \; & \; J^2  ( h^2 + \tau^2 \rt^2) \;
\end{pmatrix} , \quad
\muv = \mu \, d\xt \wedge d\yt \wedge d \zt, \quad \mu =  J h,  
\label{eqmetriccart}
\end{equation}
with $\rt^2 =  \xt^2 +\yt^2$, and we exhibit the corresponding volume form  $\muv$. 

We now make use of (\ref{eqcoordgenq}) and calculate quantities pertaining to the background flow $\Uv$: applying $(\Qv^{-1})^T$ to its components $(U,V,W)$ in (\ref{eqfullUi}) 
gives its contravariant components  as 
\begin{subequations}\label{eqUbasic1}
\begin{align}
U^1 & =   U_\cen - (S_\cen -  h^{-1} \tau W) \yt ,  
\label{eqU1basic1}\\
U^2 & = V_\cen +(S_\cen -  h^{-1} \tau W) \xt  , 
\label{eqU2basic1}\\
U^3 & = J^{-1} h^{-1} W  . 
\label{eqU3basic1}
\end{align}
\end{subequations}
Here the equations for $U^1$, $U^2$ and $U^3$ involve the full axial flow component $W$ in (\ref{eqfullU2}).
 
%
\details{This was originally set out as: 
 \begin{align*}
U^1 & =   U_\cen - P \yt ,  
\\
U^2 & = V_\cen + P \xt  , 
\\
U^3 & = J^{-1} W_\cen - J^{-1} h^{-1} (U_{\cen s} - \tau V_\cen) \xt  - J^{-1} h^{-1} (V_{\cen s} + \tau U_\cen) \yt .
\end{align*}
with 
\begin{equation*}
P  =  S_\cen -  h^{-1} \tau W .
\end{equation*}
See calculations below. 
}

The  background flow $\Uv$ generally carries a divergence, momentum and vorticity.
Its divergence is $\divv \Uv$, defined by 
\begin{equation}
d ( \Uv \ip \muv) = \muv \divv \Uv , 
\end{equation}
and in terms of components is
\begin{equation}
\divv \Uv = \mu^{-1} \bigl[ (\mu U^1)_{\xt} +   (\mu U^2)_{\yt}  +  (\mu U^3)_{\zt} \bigr]  .
\end{equation}
A short calculation using   (\ref{eqmetriccart}) and (\ref{eqUbasic1}) gives the background divergence as 
\begin{equation}
\divv \Uv = W_{\cen s}  - h^{-1} \kappa U_\cen +  h^{-1} \bigl[ - (U_{\cen s} - \tau V_\cen)_s + \tau (V_{\cen s} + \tau U_\cen)  - \kappa_s W_\cen \bigr] \xt   .
\label{eqdivvUv1} 
\end{equation}
This is the same as the result (\ref{eqmuidentity}) applied in the present $(\xt, \yt, \zt, \ttt)$ coordinate system, namely
\begin{equation}
\divv \Uv = \mu^{-1} \mu_{\ttt} = J^{-1} J_{\ttt} + h^{-1} h_{\ttt}, 
\label{eqdivvUv1a} 
\end{equation}
as may be verified with the use of (\ref{eqcurvebasisdt1}a,b) and (\ref{eqhdef}), this providing a useful check on our development.

%

The background momentum is, using (\ref{eqcoordgenq}), 
\begin{subequations}\label{eqUpsilonbasic2}
\begin{align}
\Upsilon_1  &   = U_\cen - S_\cen \yt,  \\
\Upsilon_2 &  = V_\cen +  S_\cen  \xt,  \\
\Upsilon_3  &  = J h^2 W_\cen  -  J\bigl[ h U_{\cen s} -  (1+h) \tau V_\cen\bigr] \xt   - J\bigl[h V_{\cen s} +  (1+h) \tau U_\cen\bigr]  \yt   + J \tau S_\cen   \rt^2, 
\end{align}
\end{subequations}
and, using
\begin{equation}
\Xi_1 = \partial_{\yt} \Upsilon_3 - \partial_{\zt} \Upsilon_2, \quad
\Xi_2 = \partial_{\zt} \Upsilon_1 - \partial_{\xt} \Upsilon_3, \quad
\Xi_3 = \partial_{\xt} \Upsilon_2 - \partial_{\yt} \Upsilon_1,
\end{equation}
the background vorticity is
\begin{subequations}\label{eqZetabasic1}
\begin{align}
\Xi_1  &   
= -   2 Jh (V_{\cen s}+ \tau U_\cen)  - J \bigl[ S_{\cen s} + \kappa (V_{\cen s}+ \tau U_\cen)\bigr]\xt +  2 J \tau S_\cen \yt  , 
  \\
\Xi_2  &  
=   2J  h (U_{\cen s} - \tau V_\cen)   - J \bigl[  S_{\cen s} + \kappa ( V_{\cen s}+ \tau U_\cen)\bigr] \yt  -   2 J \tau S_\cen \xt 
   + 2 J h \kappa W_\cen, \\
\Xi_3 &  
=  2 S_\cen  .
\end{align}
\end{subequations}

\details{This is also: 
\begin{align*}
\Xi_1  &   
= -    J(1+h) (V_{\cen s}+ \tau U_\cen)  - J S_{\cen s} \xt +  2 J \tau S_\cen \yt  , 
  \\
\Xi_2  &  
=   J  (1+h) (U_{\cen s} - \tau V_\cen)   - J  S_{\cen s}  \yt-   2 J \tau S_\cen \xt \notag\\
&    +  J  \kappa \bigl[ 2 h W_\cen  - (U_{\cen s} - \tau V_{\cen}) \xt - (V_{\cen s} + \tau U_{\cen}) \yt   \bigr], \\
\Xi_3 &  
=  2 S_\cen  .
\end{align*}
Note that the 1 and 2 components of (\ref{eqZetabasic1}) look a bit odd, but I think they are correct. Parts of it vanish in the slender limit in any case.
}

We now have all the machinery we need to set out the equations for vorticity in this coordinate system, in parallel with the work of many authors from \cite{Ti71} and \cite{MoSa72} onwards. We will not do so, as our aim is to move rapidly to our third and final coordinate system in \S4. However it is useful to summarise what one can do at this point, to set related work  in context, and to motivate our next steps. 

From the point of view of standard approaches, there is the inconvenience that the metric in (\ref{eqmetriccart}) is not orthogonal in general, that is for curves possessing torsion $\tau\neq0$. However the off-diagonal terms can be removed by rotating the $(\xt, \yt)$-axes about the tangent vector $\tv$ along the curve \citep{Ti71}. Once the metric is orthogonal, standard vector calculus formulae may be used if the equations of fluid motion are expressed using div, grad and curl; see for example the book \cite{TiKlKn07} and references therein. 

The next part of a standard approach is to assume a slender vortex, so that the scale transverse to the curve $\Cc$, say the radius $a(z,t)$ of the vortex, satisfies $a\kappa \equiv \eps  \ll1$, where we recall that $\kappa(z,t)$ is the filament curvature. We should note that we have not used any limit of slenderness in our discussion so far, except that clearly our coordinate system will break down if it happens that $a\kappa \simeq 1$ during filament evolution. Nonetheless at some point (we will do so later) slenderness needs to be invoked, and then a vortex is typically posited as axisymmetric at leading order, so that vorticity surfaces (tangent to the vector field $\omegav$) are approximately given by surfaces of constant $\rt$ and the flow and vorticity are dominated by their components in the $\thetat$ and $\zt$ directions. At this stage perturbation methods can be deployed to express corrections to these base fields in powers of $\eps$, with the usual powerful machinery of asymptotic matching and solvability conditions \citep[see, e.g.,][]{CaTi78,FuMi91,Ma02,TiKlKn07}. 

In summary, the vortex coordinate system $(\rt, \thetat, \zt, \ttt)$  is useful because the vorticity surfaces are approximately given by constant $\rt$ for slender vortices and that discrepancy, which is vital in determining the vortex filament evolution through curvature and external flows, can be treated by perturbation theory. We will leave this approach here, but build on it, as described in the next section. 


\section{Fluid dynamics in the Lagrangian coordinate system}\label{seclagcoord}

\noindent
In this section we leave the standard approaches behind and adopt the Lagrangian coordinate system $(r, \theta, z, t)$ defined to follow the  geometry of vorticity surfaces, as used in \cite{KoCh91} for instabilities of vortex rings without axial flow, and \cite{ChGi21} for area waves on a vortex column; for a related study in dynamo theory see \cite{GiPo00}. There are also similarities with the classic papers of \cite{Fr70,Fr72}, who solved for the shape of vortex surfaces for the case of vortex rings with no axial flow. While the $(\rt, \thetat, \zt, \ttt)$ vortex coordinate system follows vortex surfaces approximately, 
we now adapt this coordinate system to give our final system $(r, \theta, z, t)$, which follows them exactly. 

The general idea is summarised in figure \ref{fig2}: we can think of a specified vortex having a very simple structure in terms of $(r,\theta,z,t)$ coordinates (top picture) with vortex lines, curves tangent to the vorticity vector field $\omegav$, lying on nested cylindrical vorticity surfaces, of constant $r$, sitting in a reference space. This vorticity field in the reference space can vary in time, but any vortex line always sits on the same cylindrical surface. In the ambient space (lower picture) where the actual fluid dynamics takes place, the vorticity surfaces are again given by $r$ constant, but now the coordinate system $(r,\theta,z,t)$ moves and flexes. The equations for vortex dynamics become in large part equations for the coordinate system; essentially this is applying the method of strained coordinates \citep{Va75}. We stress that we have made no assumption of slenderness, and the development is exact until sections 6 and 7, where we will apply perturbation theory, principally to the description of the geometry rather than the flow and vorticity fields. 


\subsection{General framework}\label{sseclagcoordgen}

\begin{figure}
\centering
\includegraphics[scale=0.5]{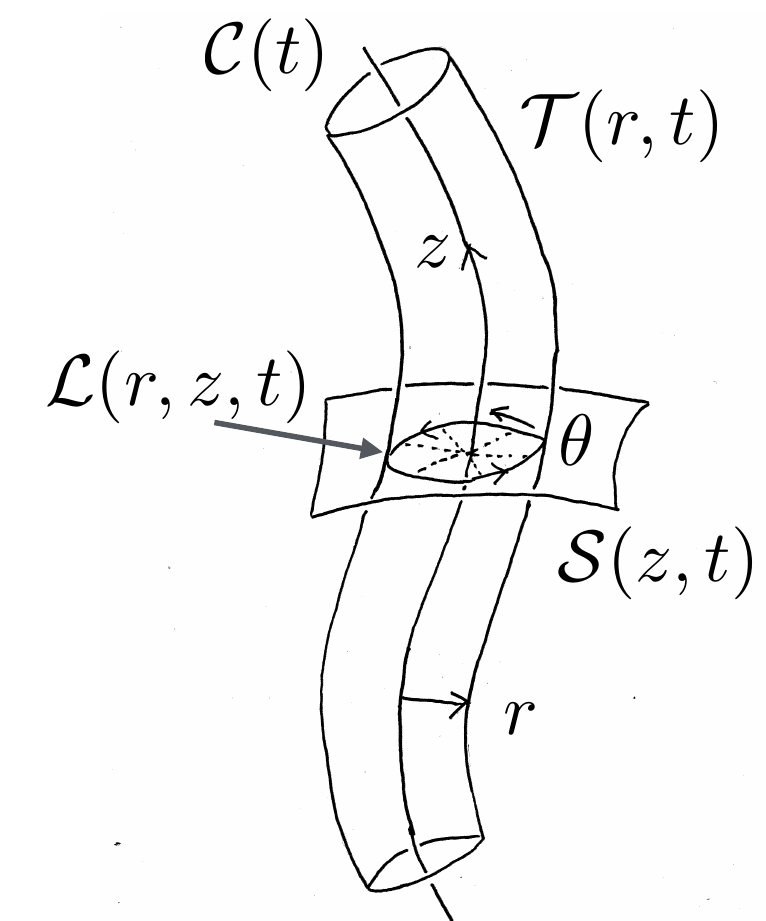}
 \caption{Coordinate geometry: at a Lagrangian station on $\Cc(t)$ given by $z$, varying the coordinates $(r, \theta)$ gives a surface $\Sc(z,t)$. Fixing $r$ and varying $(\theta,z)$ gives a tube $\Tc(r,t)$. The intersection of two of these surfaces is a curve $\Lc(r, z, t)$, on which $\theta$ is an angle coordinate.}
 \label{fig3}
\end{figure}

\noindent
We will now work in the Lagrangian coordinate system in its polar form $(r,\theta,z)$ as given in (\ref{eqrvlagcoord1cart}), where the vortex coordinates $(\xt, \yt, \zt)$ are functions of $(r, \theta, z,t)$ that we leave as arbitrary at the outset, only subject to (\ref{eqoriginset}). In other words we have from (\ref{eqrvlagcoord1cart}) that 
\begin{equation}
\xt = \xt(r,\theta,z,t), \quad
\yt = \yt(r,\theta,z,t), \quad
\zt = \zt(r,\theta,z,t) = z + q(r,\theta, z, t), 
\label{eqxtytztdef1}
\end{equation}
where we have also introduced the function $q(r,\theta,z,t)$, since the difference between $\zt$ and $z$ coordinates will be small when we later adopt the slender vortex limit, and it will be helpful to make this explicit. We require 
\begin{equation}
q(0,\theta,t) = 0, 
\end{equation}
so that on $\Cc$ the coordinates $z$ and $\zt$ coincide, as usual. The coordinate system is illustrated schematically in figure \ref{fig3}. The coordinates are pegged to the centre curve $\Cc$, where Lagrangian points are labelled by $z$. At such a Lagrangian station $z$, varying $r$ and $\theta$ sweeps out a surface $\Sc(z,t)$, and fixing $r$ within this surface gives a curve $\Lc(r,z,t)$ parameterised by the angle coordinate $\theta$ running from zero to $2\pi$. 
When $q$ is non-zero, the surfaces $\Sc(z,t)$ in figure \ref{fig3} do not coincide with the $\{\nv, \bv\}$ planes in the ambient space and so we sometimes call $q$ the \emph{tilt}.
Fixing $r$ and varying $\theta$ and $z$ gives a tubular surface $\Tc(r,t)$ on which we require vortex lines to lie, in other words there is no flux of the two-form vorticity $\xiv$ through such a surface, or equivalently $\omegav$ is tangential. 

We will specify the geometry of the vorticity distribution in Lagrangian coordinates $(r,\theta, z)$ carefully, and this will ultimately determine the evolution of the functions  $(\xt, \yt, \zt)$ of $(r, \theta, z, t)$ in (\ref{eqxtytztdef1}). In other words we are moving much of the complexity of the evolving vorticity field into complexity of the coordinate system. We use a bar to denote an average over angle $\theta$ and a prime for the fluctuating component so that, for example $v^2 = \bar{v}^2 + v^{2\prime}$. Following the approach in \cite{ChGi21},  we will require that the vorticity components satisfy
\begin{subequations}\label{eqzetalag1}
\begin{align}
\xi_1 & = 0 , \\
\xi_2 & = \xib_2(r,z,t), \quad \xi'_2 = 0 , \\ 
 \xi_3 & = \xib_3(r) , \quad\quad\ \, \;  \xi'_3 = 0,   
 \end{align}
\end{subequations}
and the relative velocity components satisfy
\begin{subequations}\label{eqvlag1}
 \begin{align}
 v^1 & = 0 , \\
 v^2 & = \vb^2(r,z,t)  , \quad v^{2\prime} = 0 , \\
 v^3 & = \vb^3(r,z,t) , \quad v^{3\prime} = 0 .
\end{align}
\end{subequations}

In our later development it is sometimes useful to conside the mean and fluctuating parts of a field obtained by piecing together its components, and so we define, for example,
\begin{equation}
\uv' = u^{1\prime} \, \partial_r + u^{2\prime} \, \partial_\theta + u^{3\prime} \,\partial_z, \quad
\uvb = \ub^1 \, \partial_r + \ub^2 \, \partial_\theta + \ub^3 \,\partial_z.
\end{equation}
This means that we could write
\begin{equation}
 \xiv = \xivb,  \quad \xiv' = 0, \quad \vv = \vvb,  \quad \vv' = 0 , \quad \xi^1 = v^1 = 0 , \quad \xi_3 = \xib_3(r), 
\end{equation}
in place of (\ref{eqzetalag1}, \ref{eqvlag1}). We will discuss the nature and consistency of these conditions in \S\ref{sseclagcoordcons} below. 


\subsection{Governing equations}\label{sseclagcoordgov}

\noindent
We now apply the conditions (\ref{eqzetalag1}, \ref{eqvlag1}) to the governing equations set out in sections \ref{ssecgeometricfull},  \ref{ssecgeometriccomp}. Inviscid fluid flow satisfies the system written in geometric form in (\ref{eqgengeom}), and in components in (\ref{eqvortcomp1}--\ref{equUvagain}).  
For coherence, we gather together all the equations that result in one place here, with some more explanation below: 
\begin{subequations}\label{eqgovlag1}
 \begin{align}
 \partial_{t}  \xib_2 & + \partial_{z} (\vb^3 \xib_2 ) =  \xib_3\,  \partial_{z} \vb^2 , \\
0 & = \partial_{\theta} \nu_{3} - \partial_{z}\nu_{2} , \\
\xib_2 & = \partial_{z}\nu_{1} - \partial_{r}\nu_{3} , \\
\xib_3 & = \partial_{r}\nu_{2} - \partial_{\theta}\nu_{1}  \\
 \nu_i & = g_{ij} u^j , \\
\partial_{r} (\mu u^1)  & + \partial_{\theta} (\mu u^2)  + \partial_{z} (\mu u^3)  = 0 , \\
u^1  = U^1, \quad u^2 & = U^2 + \vb^2, \quad u^3 = U^3 + \vb^3. 
\end{align}
\end{subequations}
The main simplification arising from (\ref{eqzetalag1}, \ref{eqvlag1}) is that the vorticity equation, in components in (\ref{eqvortcomp1}), collapses to just one equation (\ref{eqgovlag1}a) for the azimuthal component $\xib_2$. This represents the winding up of the fixed component $\xib_3(r)$ by the differential rotation of $\vb_2$ to produce $\xib_2$, which then undergoes axial transport by $\vb^3$. Other aspects of vorticity evolution are taken up by the coordinate system. Equations (\ref{eqgovlag1}b--d) relate the momentum components $\nu_i$ to the simple imposed vorticity structure, from (\ref{eqzetaagain}). The needed geometric information is in the metric and volume form in (\ref{eqgovlag1}e, f) and in the background flow in (\ref{eqgovlag1}g), taken from (\ref{eqnuagain}--\ref{equUvagain}), and we turn to these next.


\subsection{Metric and background flow}\label{sseclagcoordback}

\noindent
As we are now using the Lagrangian coordinates $(r, \theta, z)$ in place of vortex coordinates $(\xt, \yt, \zt)$ the background flow $\Uv$ changes from that defined earlier. This then results in changes to $\Upsilonv$,  $\Xiv$, $\vv$, $\piv$ and $\varpiv$ in (\ref{eqschema}), as these are objects defined relative to whatever moving coordinate system we use. This situation can contrasted with that for $\gv$, $\muv$, $\uv$, $\nuv$, $\xiv$, which are defined as geometric objects independently of our choice of coordinate system, albeit that their components will change when we refer to the new bases $\{ \partial_r, \partial_\theta, \partial_z\}$ and $\{ dr, d\theta, dz\}$. We commence the  determination of these objects in component form, building on our earlier calculations (and recycling some notation). 

Differentiating $\rv$ in (\ref{eqrvlagcoord1cart}) with respect to time, we find the components of the background flow $\Uv$ with respect to the $\{\nv, \bv, \tv\}$ basis are
\begin{subequations}\label{eqUVWlag1}
\begin{align}
 & U  = U_\cen   - S_\cen \yt +  \xt_t   -  J \tau  q_t \yt ,  
  \label{eqfullUgen00a} \\
 & V  = V_\cen  + S_\cen \xt  +  \yt_t    + J \tau  q_t \xt ,  \\
 & W  = W_\cen h   - ( U_{\cen\arcl} - \tau V_\cen) \xt - (V_{\cen\arcl} + \tau U_\cen)\yt   + J h q_t  .
 \label{eqfullUgen02a}
\end{align}
\end{subequations}
The matrix $\Qv$ is given by differentiating (\ref{eqrvlagcoord1cart}) with respect to the spatial coordinates to obtain, using (\ref{eqgendrq}), 
\begin{equation}
\Qv =
\begin{pmatrix}
\xt_r & \yt_r & q_r \\
\xt_\theta & \yt_\theta  & q_\theta \\
\;\xt_z\;  & \;\yt_z \; & \;1 + q_z \;
\end{pmatrix} 
\begin{pmatrix}
1 & 0 & 0 \\
0 & 1 & 0 \\
- J \tau \yt  & J \tau \xt  & Jh
\end{pmatrix} .
\label{eqQvfull}
\end{equation}
The product of matrices here arises from the chain rule, adding a layer of complexity to our earlier calculations. Space and time derivatives of the unspecified functions $(\xt, \yt, q)$ of $(r,\theta,z,t)$ in (\ref{eqxtytztdef1}) appear here and in (\ref{eqUVWlag1}), confirming that we are at the start of generating PDEs for the coordinate system. 

Going further is straightforward in principle, to calculate $\gv$ and $\muv$ with respect to these coordinates from (\ref{eqgengqq}) and to continue to determine the background flow quantities $(U^1, U^2, U^3)$, $\divv \Uv$, $\Upsilonv$ and $\Xiv$, analogously to calculations in \S\ref{ssecvortcoordback}. We will not do so as the calculations rapidly become unwieldy and uninformative; nonetheless it is clear how these quantities are defined. To make sensible progress we will need to take the limit of a slender vortex and apply perturbation theory, which we take up in section 5.

\subsection{Consistency}\label{sseclagcoordcons}

\noindent
We have applied many conditions in (\ref{eqzetalag1}, \ref{eqvlag1}), and before we move to the slender vortex limit, it is worth pausing to consider issues of consistency and gauge freedom. Let us freeze time and discuss the frozen geometry, referring to figure \ref{fig3}.

Condition (\ref{eqzetalag1}a) means that the tubular surfaces $\Tc(r)$ of constant $r$ are tangential to vortex lines, or equivalently that there is no flux of the two-form vorticity $\xiv$ through such surfaces, so that its component $\xi_1=0$. Condition (\ref{eqvlag1}a) then ensures that these surfaces stay at constant coordinate values of $r$ as there is no relative velocity $v^1$ across them. The flow components $v^2$ and $v^3$ then lie in these surfaces of constant $r$, and the coordinate $\theta$ is arranged so that the angular velocity $v^2$ around each curve $\Lc(r,z)$ in figure \ref{fig2}  is constant, $v^2 = \vb^2$, which is the condition (\ref{eqvlag1}b). This is simply the usual definition of an angle coordinate in the sense of action--angle coordinates, and can clearly be achieved by flexing the surfaces of constant $\theta$ coordinate. Similarly by flexing the surfaces of constant $z$, we can achieve that the axial velocity is constant, $v^3 = \vb^3$, on curves $\Lc(r,z)$, giving condition (\ref{eqvlag1}c). 


This leaves behind conditions (\ref{eqzetalag1}b, c) to discuss. Let us check the consequences of applying only the conditions (\ref{eqzetalag1}a, \ref{eqvlag1}a--c) that we have explained so far, but not (\ref{eqzetalag1}b, c) and allow general $\xi_2(r, \theta, z, t)$ and $\xi_3(r, \theta, z, t)$. In this case, first note that (\ref{eqdxizero}) becomes
\begin{equation}
\partial_\theta \xi_2 + \partial_z \xi_3 = 0 , 
\label{eqdxizero1}
\end{equation}
and on taking the $\theta$-average we have that $\partial_z \xib_3 = 0$ so $\xib_3= \xib_3(r,t)$ only. The remaining, fluctuating components are related by 
\begin{equation}
\partial_\theta \xi'_2 + \partial_z \xi'_3 = 0 . 
\label{eqdxizerop}
\end{equation}

Taking this information, that $\xib_3$ is independent of $z$ as well as $\theta$, and making use of (\ref{eqzetalag1}a, \ref{eqvlag1}a--c), we simplify the three components of the vorticity equation in (\ref{eqvortcomp1}). It is evident that (\ref{eqvortcomp1}a) is satisfied trivially, while the remaining two components are 
\begin{subequations}\label{eqvortcomp1b}
\begin{align}
 & \bigl(\partial_{t} + \vb^2 \partial_{\theta} + \vb^3 \partial_{z}\bigr) \xi_2 -   \xi_3 \partial_{z} \vb^2 
 +  \xi_2  \partial_z \vb^3 = 0 , \\ 
& \bigl(\partial_{t} +   \vb^2 \partial_{\theta} + \vb^3 \partial_{z}\bigr) \xi_3   = 0  .
\end{align}
\end{subequations}
If we take the $\theta$-average of (\ref{eqvortcomp1b}a) we obtain (\ref{eqgovlag1}a).  The average of (\ref{eqvortcomp1b}b) gives $\partial_t \xib_3 = 0$ and so we have established that $\xib_3 = \xib_3(r)$: axial vorticity is a fixed distribution that does not change during time evolution of the vortex, as indicated in (\ref{eqzetalag1}c). 
The fluctuating components of (\ref{eqvortcomp1b}) are 
\begin{subequations}\label{eqvortcomp1c}
\begin{align}
 & \bigl(\partial_{t} + \vb^2 \partial_{\theta} + \vb^3 \partial_{z}\bigr) \xi'_2 -   \xi'_3 \partial_{z} \vb^2 
 +  \xi'_2  \partial_z \vb^3 = 0 , \\ 
& \bigl(\partial_{t} +   \vb^2 \partial_{\theta} + \vb^3 \partial_{z}\bigr) \xi'_3   = 0  , 
\end{align}
\end{subequations}
these two equations being equivalent by virtue of the relation (\ref{eqdxizerop}). 

We will now set $\xi'_2$ and $\xi'_3$ identically zero, and there are at least three interlinked reasons to do so. First, these equations indicate that if they are zero initially, they remain zero for all time, so for simplicity it makes sense to do so, to minimise the complexity of the system we are treating. We are embedding, in the ambient space, one or more vortices in order to study their dynamics; we are free to choose their structure, and there is no point in making these more complicated than is necessary, without some particular reason. 

Secondly, suppose these fields are non-zero initially, then $\xi'_3$ is transported as a passive scalar by the flow with components $\vb^2$, $\vb^3$ in (\ref{eqvortcomp1c}b). Any variation in angular velocity $\vb^2$ with radius will lead to rapid shredding of the field $\xi'_3$, leading it to become of small scale and tend to be passive in the dynamics of the filament. Likewise any variation in angular velocity with $z$ will lead to shredding. Thus in the evolution, any initial fields $\xi'_2$, $\xi'_3$ will tend to be driven to fine scales where they are unimportant in the dynamics. 

The third reason we offer is that when we introduce slender vortex scalings below we will take $\vb_2 = O(1)$, $\partial_t$, $\partial_z =O(\eps)$ for $\eps \ll1$. These scalings give $\vb^2 \partial_\theta \xi'_3 =O(\eps)$ from (\ref{eqvortcomp1c}b) and so enforce $\xi'_3=0$ at leading order, that is at $O(1)$; then necessarily $\xi'_2 = 0$ to $O(\eps)$ from (\ref{eqdxizerop}). Another view of this is that for studying slender vortex dynamics we want to remove processes that happen on the $O(1)$ vortex turnover time scale (certainly shredding, but also potentially waves and instabilities), to concentrate on slower evolution of a filament, so we cannot allow fast evolution of these fluctuating fields. This enforces the \emph{compatibility conditions} of zero $\xi'_2$, $\xi'_3$, as discussed in, for example, \cite{TiKlKn07}.

For all these reasons we will take $\xi'_2$, $\xi'_3$ to be identically zero initially, so they remain so for all time, which means that the remaining constraints on the vorticity we impose are given in  (\ref{eqzetalag1}b, c). We also note that the gauge freedom in (\ref{eqgauge1}) now extends to replacing $(z,\theta)$  by $(\zh, \thetah)$ with 
\begin{equation}
\thetah = \theta + 
\vartheta (r,z,t), \quad
\zh =  \zh (r,z,t), 
\label{eqguageagain}
\end{equation}
in other words a reparameterisation of the origin of the angle coordinate $\theta$ and the value of $z$, on each curve $\Lc(r,z,t)$. This changes preserves the requirements (\ref{eqzetalag1}, \ref{eqvlag1}). As we will not use this gauge flexibility in any explicit way, we do not set out further details here. 


\section{Slender vortex limit}\label{secslender}

\noindent
Our development so far has been exact, though we stopped short in our calculations in section \ref{sseclagcoordback} because the general framework becomes unwieldy at this point. To make some progress, in this section we introduce the slender vortex scalings, which allow us to write down quantities as expansions in a slenderness parameter $\eps$. 


\subsection{Definition of slender vortex scalings} \label{ssecslenderdef}

\noindent
 In the Lagrangian coordinates (or reference space) the vortex will be given by $0\leq \theta \leq 2 \pi$, $0\leq z \leq 2 \pi \ell$ with periodic boundary conditions, together with $0\leq r \leq a$. Here $a$ and $\ell$ are  constants and the slender vortex limit is $\ell/a \gg1$. We will choose our basic length scale and velocity scale based on the vortex cross section and flow there, and so we will take 
\begin{equation}
a = O(1), \quad \ell \gg 1, \quad \eps \equiv a/\ell \ll1. 
\end{equation}
%
It is convenient to introduce $\eps$ here as our small expansion parameter. With this we have for the coordinates
\begin{subequations}\label{eqslender1}
\begin{align}
& x,\, y,\, r,\, \theta,\, \xt,\, \yt,\, \rt,\, \thetat = O(1) , \quad  \partial_x, \, \partial_y,\,  \partial_r,\, \partial_\theta , \,
\partial_{\xt}, \, \partial_{\yt},\,  \partial_{\rt},\, \partial_{\thetat}= O(1), \\
& z, \,\zt,\, s = O(\eps^{-1}), \quad \partial_z,\, \partial_{\zt},\,\partial_s = O(\eps) , \quad  
t = \ttt = O(\eps^{-1}), \quad \partial_t, \,\partial_{\ttt} = O(\eps). 
\end{align}
\end{subequations}
Note that we assume time evolution with respect to $t = \ttt$ on a slow time scale, slower than the $O(1)$ vortex turnover time scale. There is the freedom to introduce further, slower time scales depending on what is being modelled; see, e.g., \cite{Ma02} and \cite{TiKlKn07} for discussion of time scales and processes. 

The $\zt(r,\theta,z,t)$ map requires a little care;  we will set the tilt $q$ in (\ref{eqxtytztdef1}) to be of order $\eps$ so that we have 
\begin{equation}
  q = O(\eps),  \quad
q_r = O(\eps), \quad
q_\theta = O(\eps), \quad
q_z = O(\eps^2), \quad
q_t = O(\eps^2). 
\end{equation}
This means that the flexing and tilting of the surface $\Sc(z,t)$ about the $\{\nv, \bv\}$ plane, at any $z$-station in the ambient space in figure \ref{fig3}, is through small angles of order $\eps$.

For the properties and evolution of the centre line curve $\Cc$ we take
\begin{equation}
J = O(1), \quad \kappa, \,\tau, \,S_\cen = O(\eps) , \quad
U_\cen, \,V_\cen, \,W_\cen =O(1),
\end{equation}
in (\ref{eqcurvebasisdt1}) and note that no terms drop out of these equations by virtue of these scalings. We will also need to assume that $\kappa$ does not get too small, in other words avoiding zero crossings, since the quantity $S_\cen$ in (\ref{eqS0def}) could blow up in this context, and the system become disordered. The discontinuous behaviour of the $\{\nv, \bv, \tv\}$ basis in this case is well known \citep[see, e.g.][]{MoRi92}. In other words we will assume that $\kappa$ is bounded both above and below by $O(\eps)$ constants, and will not remark on this further. 

For the extension of the centre line motion to the background flow we then have  
\begin{equation}
U, \,V, \,W= O(1); 
\end{equation}
see (\ref{eqUVWlag1}). 
We also require for the non-zero components of the velocity field, momentum and vorticity, that
\begin{subequations}
\begin{align}
& u^1, \,u^2, \,u^3, \,\vb^2, \, \vb^3 = O(1), \\
& \nu_1,\, \nu_2, \,\nu_3 = O(1), \\
& \xib_2, \,\xib_3 = O(1). 
\end{align}
\end{subequations}

Note that our aim  here is to keep flow quantities as large as they can be, to allow the most general development of our geometric framework. However when specialised to some particular system, a given geometry in the ambient space, many quantities may be zero by symmetry, or smaller than in the estimates above. As an example we will shortly consider a steadily propagating vortex ring for which $\partial_t$, $\partial_z = 0$, $U_\cen$, $W_\cen=0$ and $V_\cen=O(\eps)$, which leads to many simplifications. However in the theoretical development, our goal here is to maintain the greatest generality.

We will keep all terms of $O(1)$ and $O(\eps)$ in the  development that follows and we find it convenient to write `$+\cdots$' to denote further terms that are $O(\eps^2)$. Other error estimates will be given explicitly.


\subsection{Metric and background flow}\label{ssecslendermetric}

\noindent
Let us pick up the discussion from around (\ref{eqUVWlag1}); we have in the slender vortex limit:
\begin{subequations}
\label{eqfullUgen0b}
\begin{align}
 & U  = U_\cen   - S_\cen \yt +  \xt_t   + O(\eps^3)   , 
 \label{eqfullUgen00b} \\
 & V  = V_\cen  + S_\cen \xt  +  \yt_t    + O(\eps^3)  ,  \\
 & W  = W_\cen h   - ( U_{\cen\arcl} - \tau V_\cen) \xt - (V_{\cen\arcl} + \tau U_\cen)\yt   + \cdots  .
 \label{eqfullUgen02b}
\end{align}
\end{subequations}
This is the previous background flow (\ref{eqfullUi}) in the vortex coordinate system plus a new piece from the additional flexing of the Lagrangian coordinate system, which it is convenient to call
\begin{equation}
\Uv_{\flex} = \xt_t \nv + \yt_t \bv + \cdots. 
\label{eqUflexdef}
\end{equation}
We have further flexing terms  in (\ref{eqUVWlag1}) but these are at $O(\eps^2)$ or smaller and so neglected here. 
For the matrix $\Qv$ in (\ref{eqQvfull}) we have,
\begin{subequations}
\label{eqQvslender}
\begin{align}
\Qv  =
&\begin{pmatrix}
\xt_r & \yt_r & Jq_r \\
\xt_\theta & \yt_\theta  & Jq_\theta \\
\;\xt_z  - J \tau \yt \;  & \;\yt_z + J \tau \xt\; & \;Jh \;
\end{pmatrix}  + \cdots, 
\\
(\Qv^{-1})^T  =
\alpha^{-1} &
\begin{pmatrix}
\;\yt_\theta \;& \;- \xt_\theta \; &\; J^{-1} (\xt_\theta \yt_z - \xt_z \yt_\theta) + \tau \rt \rt_\theta  \; \\
- \yt_r & \xt_r  & \; J^{-1} (\xt_z \yt_r - \xt_r \yt_z) - \tau \rt \rt_r   \;\\
\; \yt_r q_\theta - \yt_\theta q_r  \; & \;\xt_\theta q_r - \xt_r q_\theta  \;& \; J^{-1} h^{-1} \alpha \;
\end{pmatrix}  + \cdots , 
\end{align}
\end{subequations}
with $\alpha = \xt_r \yt_\theta - \xt_\theta \yt_r$ here and in (\ref{eqalphaslender}) below.
Note that in view of (\ref{eqhdef}) we replace $h$ by one whenever we can using $h = 1 + O(\eps)$; also we have $h_t$, $h_z = O(\eps^2)$. From (\ref{eqgengqq}) applied in the present context we find the components of the metric 
\begin{equation}\label{eqmetricslender}
(g_{ij}) =  
\begin{pmatrix}
\xt_r^2 + \yt_r^2 &\; \xt_r \xt_\theta + \yt_r \yt_\theta \; & \xt_r \xt_z + \yt_r \yt_z  + J \tau ( \xt \yt_r - \yt \xt_r) + J^2 q_r \\
\otimes & \xt_\theta^2 + \yt_\theta^2 &   \xt_\theta \xt_z + \yt_\theta \yt_z + J \tau ( \xt \yt_\theta - \yt \xt_\theta) + J^2 q_\theta \\
\otimes & \otimes& J^2h^2
 \end{pmatrix}
 +\cdots , 
 \end{equation}
(omitting to write three entries as it is symmetric) together with the volume form
\begin{equation}
\muv = \mu \, dr\wedge d\theta \wedge dz, \quad \mu =     J  h \alpha + \cdots , 
\label{eqmuslender}
\end{equation}
where we define $\alpha$ via the area element $\alphav$ in the $(r,\theta)$ plane at fixed $z$ by 
\begin{equation}
\alphav = \alpha \, dr\wedge d\theta, \quad \alpha = \xt_r \yt_\theta - \xt_\theta \yt_r .
\label{eqalphaslender}
\end{equation}
Equivalently $\alpha^2$ is the determinant of the first $2\times2$ block of the metric,
\begin{equation}
\alpha^2 = g_{11}g_{22} - g_{12} g_{12}. 
\label{eqalphaslender1}
\end{equation}
We now have all the ingredients we need, namely $\Qv$, $\gv$ and $\muv$,  to calculate the contravariant components $U^i$ of the background flow $\Uv$ in (\ref{eqfullUgen0b}), and the corresponding momentum $\nuv$ and vorticity $\Xiv$, as we did for the vortex coordinate system in \S\ref{ssecvortcoordback}. We will continue the general development in section \ref{secgeneralslender}.


\section{Vortex ring structure and motion}\label{secring}

\noindent
At this point in the paper we have developed the geometrical framework, in particular the use of the Lagrangian coordinate system, and  discussed the scalings appropriate to a slender vortex. The centre line equations (\ref{eqcurvebasisdt1}) remain unchanged under these scalings, and in the previous  \S\ref{ssecslendermetric} we have put in place what we need to calculate all relevant background flow quantities (see \S\ref{secbackgroundslender} below). As the general development becomes more complicated at this point, it is helpful to pause and apply what we know so far to the classic example of a slender, steadily propagating vortex ring. 
We consider a vortex ring of circumference  $2\pi \ell$, curvature $\kappa = \ell^{-1}$ and minor radius $a$, with $\eps = a / \ell \ll1$. 
We will calculate the velocity of the ring as a function of its structure, a classic calculation set out in, for example \cite{Sa92} making use of arguments involving impulse, and by \cite{WiBlZa71} using a perturbation series in the equivalent of the vortex coordinate system. Our goal here is to provide a purely vorticity based view of vortex ring motion, in a Lagrangian coordinate system, and so insights into how our general framework fits together. We then resume the general development in \S\ref{secfullgeneral}. 


\subsection{Vortex column}\label{ssecringstraight}

\noindent
A vortex ring tends to a straight, columnar vortex in the slender limit, that is as $\ell\to\infty$ for fixed $a$ with $a / \ell = \eps \to0$, $\kappa=\ell^{-1} = O(\eps)$, $a=O(1)$. To employ perturbation theory we first need to define a basic state of a straight vortex column with $\eps=\kappa=0$. We set
\begin{subequations}
\begin{align}
\rt  &= r, \quad \thetat = \theta, \quad \zt = z, 
\label{eqstraight1}
\\
\partial_t ,\, \partial_\theta , \,\partial_z,\, \partial_s = 0, \quad q &= 0 , \quad J=1, \quad   \tau ,\,  \kappa,  \,S_\cen = 0 , \quad U_\cen,\, V_\cen,\,W_\cen = 0  , 
\label{eqstraight2}
\end{align}
\end{subequations}
and recall that $(\xt, \yt)$ are always linked to $(\rt, \thetat)$ via (\ref{eqxtytrtthetat}). There is no background flow. The non-zero metric coefficients are just $g_{11}=1$, $g_{22} = r^2$ and $g_{33} = 1$, and we have $\mu = r$. 

The vortex is governed by (\ref{eqgovlag1}) with (\ref{eqgovlag1}a, b) automatically satisfied. We are free to specify the two components $\xib_2(r)$ and $\xib_3(r)$ of the vorticity distribution, and then the vorticity--momentum link (\ref{eqgovlag1}c, d) gives the momenta $\nu_2 = \nub_2$ and $\nu_3 = \nub_3$ via 
\begin{equation}
\xib_2 = - \partial_r \nub_3, \quad
\xib_3 =  \partial_r \nub_2. 
\label{eqxi23relate}
\end{equation}
From the momentum--flow link in (\ref{eqgovlag1}e) we obtain the flow components 
\begin{equation}
\nub_3 =  \vb^3 , \quad \nub_2 = r^2 \vb_2 .
\label{eqnu23relate}
\end{equation}
Equation (\ref{eqgovlag1}f) is satisfied and (\ref{eqgovlag1}g) is simply
\begin{equation}
u^1 = 0 , \quad u^2 = \vb^2, \quad u^3 = \vb^3. 
\end{equation}

We can keep a general basic state vorticity and flow distribution, related by (\ref{eqxi23relate}, \ref{eqnu23relate}),  or illustrate with a particular choice. For example the following flow has constant axial vorticity per unit area and a parabolic axial velocity profile: 
\begin{subequations}
\begin{align}
\xib_{3} = \begin{cases} 2 \Omega r, & \\ 0, & \\ \end{cases}
\quad
\nub_{2}  = \begin{cases}  \Omega r^2, & \\ \Omega a^2, & \\ \end{cases}
\quad
\vb^{2} & = \begin{cases}  \Omega, & \\ \Omega a^2 / r^2 , & \\ \end{cases}
\quad
\psib = \begin{cases} - \tfrac{1}{2} \Omega r^2, &  \\  - \tfrac{1}{2} \Omega a^2  - \Omega a^2 \log (r/a) ,& \\ \end{cases}
\label{eqbasicstatezeta3}
\\
\xib_{2} = \begin{cases} \chi  r ,& \\ 0, & \\ \end{cases}
\quad
\nub_{3} & = \vb^3 =   \begin{cases} \tfrac{1}{2} \chi (a^2-r^2),  \\ 0 , &  \end{cases}
 \label{eqbasicstatezeta2}
\end{align}%
\label{eqbasicstatezeta}%
\end{subequations}%
where here and below the upper expression is taken for $r\leq a$ and the lower for $r> a$. For convenience we have also defined a stream function $\psi = \psib(r)$, with 
\begin{equation}
\vb^2 = - r^{-1} \partial_r \psib . 
\label{eqvb2psib}
\end{equation}
This all provides a $\kappa = \eps = 0$ basic state %
which we now perturb, by bending this gently into a vortex ring.


\subsection{Governing equations}\label{ssecringgov}

\noindent
We revisit the equations for $\kappa$ non-zero but small and positive. Both $\eps$ and $\kappa$ are of the same order since we take $a = O(1)$, $\eps = a / \ell = a \kappa$. Thus we can equally take our expansion parameter as $ \kappa \ll1$, and think of the small bending of the original vortex column as driving the asymptotic expansion and giving $O(\kappa)$ corrections to the basic state in \S\ref{ssecringstraight}. Our goal is to calculate all corrections of this order to the above basic state, out of which will come the propagation velocity of the vortex ring. 

We anticipate the structure of the solution by assuming a ring propagating steadily in the binormal direction.  We relax (\ref{eqstraight2}) to 
\begin{equation}
\label{eqcurved2}
 \partial_t  , \,\partial_z ,\, \partial_s = 0,  \quad J=1, \quad \tau, \,  S_\cen=0, \quad  U_\cen, \, W_\cen = 0 .  
\end{equation}
In place of (\ref{eqstraight1}) we will allow the more general functional forms $\rt (r, \theta)$, $\thetat (r, \theta)$, $\zt (r, \theta)$, and also set 
\begin{equation}
0< \kappa =\text{const.}  = O(\eps), \quad  V_\cen  =\text{const.}  = O(\eps).
\end{equation}
%

The filament centreline equations (\ref{eqcurvebasisdt1}) hold trivially,  and we turn to the governing system in  (\ref{eqgovlag1}). The vorticity equation (\ref{eqgovlag1}a) is satisfied and the remaining  equations become
\begin{subequations}\label{eqgovlag2}
 \begin{align}
0 & = \partial_{\theta} \nu_{3} , \\
\xib_2 & = - \partial_{r}\nu_{3} , \\
\xib_3 & = \partial_{r}\nu_{2} - \partial_{\theta}\nu_{1} ,  \\
 \nu_i & = g_{ij} u^j , \\
\partial_{r} (\mu u^1)  & + \partial_{\theta} (\mu u^2) = 0 , \\
u^1  = U^1, \quad u^2 & = U^2 + \vb^2, \quad u^3 = U^3 + \vb^3. 
\end{align}
\end{subequations}
Equation (\ref{eqgovlag2}a) tells us that $\nu'_3 = 0$ and, together with the averages of  (\ref{eqgovlag2}b,c), we have 
\begin{equation}
\xib_3 = \partial_{r}\nub_{2} , \quad \xib_2 = - \partial_{r}\nub_{3} ,   \quad \nu_3 = \nub_3, \quad \nu'_3 = 0 . 
\label{eqxilots1}
\end{equation}
The fluctuating part of  (\ref{eqgovlag2}c) gives the key relationship 
\begin{equation}
0 =  \partial_{r}\nu'_{2} - \partial_{\theta}\nu'_{1}, 
\label{eqnu12solve1}
\end{equation}
while equations  (\ref{eqgovlag2}a--c) do not restrict $\nub_1$, which is simply a diagnostic that could be calculated from  (\ref{eqgovlag2}d) and turns out to be zero (to the order we are working) in (\ref{eqnub123approx}). 
As for the column vortex in \S\ref{ssecringstraight},  we can make a free choice of $\xib_3(r)$ and $\xib_2(r)$ and then these fields give the core structure in terms of vorticity together with the mean momenta, $\nub_2$, $\nu_3 = \nub_3$ in (\ref{eqxilots1}). 


\subsection{Metric and background flow}\label{ssecringmetric}

\noindent
Perturbing the column vortex in \S\ref{ssecringstraight}, we replace (\ref{eqstraight1}) by nearly circular streamlines with 
\begin{equation}
\rt = r [ 1 + f(r)\cos \theta]  + \cdots , \quad \thetat = \theta + g(r)\sin \theta  + \cdots , \quad \zt = z+ q(r) \sin \theta + \cdots, 
\label{eqcoordtcoord}
\end{equation}
where $f$, $g$ and $q$ are functions of order $\eps$. Here we are anticipating the $\theta$-dependence to minimise complexity; the non-zero $\kappa$ excites terms in $\sin \theta$ and $\cos\theta$ only, to the order at which we are working. We have replaced the tilt $q(r,\theta)$ by $q(r) \sin \theta$ for convenience. Recall that the notation `$+ \cdots$' is short for `${} + O(\eps^2)$' and that $\kappa = O(\eps)$. 


A short calculation gives, from (\ref{eqmetricslender}--\ref{eqalphaslender}), also (\ref{eqxtytrtthetat}), 
\begin{subequations}
\begin{align}
(g_{ij}) & =  
\begin{pmatrix}
1 + ( 2r f_r + 2f) \cos \theta &\; (- rf + r^2 g_r)\sin \theta \; &  q_r  \sin \theta\\
\otimes & r^2[ 1 + 2(f +g )\cos \theta  ] &     q \cos \theta \\
\otimes & \otimes& 1 - 2 \kappa r \cos \theta 
 \end{pmatrix}
 +\cdots , 
\label{eqg1}
\\
\alpha & = 
r [ 1 + ( r f_r + 2f + g)\cos \theta ]  + \cdots , \\
 \mu  & = \alpha h = r [  1 + (rf_r + 2 f + g - \kappa r )\cos \theta]   + \cdots, 
 \label{eqmu1}
\end{align}
\end{subequations}
and similarly we have $\Qv$ and $\Qv^T$ from (\ref{eqQvslender}); we omit the details. 
The background flow is simply 
\begin{equation*}
U = 0, \quad V = V_\cen, \quad W=0, 
\end{equation*}
approximately from (\ref{eqfullUgen0b}) or exactly from (\ref{eqUVWlag1}), using $S_\cen=0$, and we have from (\ref{eqcoordgenq}) and (\ref{eqQvslender}) the components for this and the momentum as 
\begin{subequations}
\label{eqUUpsring}
 \begin{align}
 \!\!\!
 U^1 & = 
 V_\cen \sin \theta +\cdots ,   \quad 
 U^2 =     
 V_\cen  r^{-1} \cos \theta  + \cdots ,\quad 
 U^3 =  0 + \cdots ,  \\
\!\!\!
\Upsilon_1 & =   
V_\cen \sin \theta    + \cdots  , \quad 
\Upsilon_2 =  
V_\cen r \cos \theta  + \cdots  , \quad 
\Upsilon_3 = 0 + \cdots .
\end{align}
\end{subequations}
Naturally the background flow does not carry a divergence or vorticity,  $\divv \Uv = 0$, $\Xiv = 0$,
being uniform in the ambient space.

\subsection{Vortex ring structure}\label{ssecringstruct}

\noindent
We now start to solve the governing equations, working from (\ref{eqgovlag2}d--f, \ref{eqxilots1}, \ref{eqnu12solve1}). Since $\Upsilon_i = g_{ij} U^j$ gives the background momentum in (\ref{eqschema}, \ref{eqschemarels}), we can subtract this from both sides of (\ref{eqgovlag2}d), and so rewrite this as 
\begin{equation}
\nu_i - \Upsilon_i = g_{ij} v^j . 
\end{equation}
Putting in $\Upsilonv$ from (\ref{eqUUpsring}b), making use of (\ref{eqvlag1}, \ref{eqxilots1}, \ref{eqg1}) and writing out the components yields 
\begin{subequations}
\begin{align}
\nu_1 - V_\cen \sin \theta  &  = [ (-rf + r^2 g_r) \vb^2 + q_r \vb^3]  \sin \theta + \cdots, \\
\nu_2 - V_\cen r \cos \theta   & = r^2 \vb^2 + [ 2r^2(f+ g) \vb^2 + q\vb^3] \cos \theta + \cdots, \\
\nu_3 = \nub_3 & = \vb^3 + (q \vb^2 - 2\kappa r \vb^3 ) \cos \theta + \cdots.
\end{align}
\label{eqringnurel}
\end{subequations}
If we take the mean of these equations we gain the relationships
\begin{equation}
\nub_1 = 0 + \cdots, \quad \nub_2 = r^2 \vb_2+ \cdots, \quad \nub_3 = \vb^3+ \cdots.
\label{eqnub123approx}
\end{equation}
The fluctuating part of (\ref{eqringnurel}c) fixes the tilt $q$ in terms of other quantities, 
\begin{equation}
q =  2 \kappa r \vb^3 / \vb^2, 
\label{eqringqrel}
\end{equation}
and expresses how curvature leads to tilting of the surfaces $\Sc$ in, say, figure \ref{fig3}, as noted by \cite{WiBlZa71}. 

The fluctuating parts of (\ref{eqringnurel}a,b) give 
\begin{subequations}
\begin{align}
\nu'_1  &  = [ V_\cen + (-rf + r^2 g_r) \vb^2 + q_r \vb^3]  \sin \theta + \cdots, \\
\nu'_2     & = [ r V_\cen + 2r^2(f+ g) \vb^2 + q\vb^3] \cos \theta + \cdots,
\end{align}
\label{eqringnurelfluc}
\end{subequations}
and we can substitute these into the condition (\ref{eqnu12solve1}), which we recall imposes zero fluctuating axial vorticity. The ring velocity $V_\cen$ drops out to leave 
\begin{equation}
2r f_r + (2 \sigmab + 1) f + r g_{r} + 2 \sigmab g  + \rhob \kappa r  = 0 , 
\label{eqringODE1}
\end{equation}
where we abbreviate
\begin{equation}
\sigmab(r) =  \frac{( r^2 \vb^2)_r }{ r \vb^2 }\, , \quad
\rhob (r) = \frac{ [ (\vb^3)^2]_r } { r (\vb^2)^2}\, . 
\label{eqdefsimgabrhob}
\end{equation}
and we have used $q$ from (\ref{eqringqrel}). This is the first ODE we need, to determine $f$ and $g$ and obtain the vortex ring structure, and originates from the requirement that $\xi'_3 = 0$ in (\ref{eqzetalag1}c). 

For a second ODE we use the one equation we have not yet applied, namely the incompressibility condition (\ref{eqgovlag2}e). We check that $\divv \Uv = \mu^{-1} \mu_t =0 $, either exactly from (\ref{eqmuidentity}), or here by calculating, to this order, 
\begin{equation}
\divv \Uv = \partial_r (\mu U^1) + \partial_\theta(\mu U^2) = \partial_r (r V_\cen \sin \theta ) + \partial_\theta(V_\cen \cos\theta) + \cdots = 0 + \cdots.  
\end{equation}
Then we have, as $\uv = \Uv + \vv$ in (\ref{eqgovlag2}f), that $\divv \uv =0$ in (\ref{eqgovlag2}e) becomes 
\begin{equation}
0 = \divv \vv  =  \divv \vvb  = \mu^{-1} \partial_\theta(\mu \vb^2) = \mu^{-1} \vb^2 \partial_\theta\mu = 0 , 
\end{equation}
which gives  $\partial_\theta \mu = 0$ (noting we need a non-zero flow $\vb^2$ to have a vortex filament in the first place), and so 
\begin{equation}
\mu' = 0 . 
\end{equation}
This result, which is exact, applied with (\ref{eqmu1}) results in our second ODE:
\begin{equation}
rf_r + 2 f + g - \kappa r = 0 . 
\label{eqringODE2}
\end{equation}

Eliminating $g$ from the two ODEs (\ref{eqringODE1}) and (\ref{eqringODE2}) gives 
\begin{equation}
r^2 f_{rr} + (2 \sigmab+1) r f_r + (2 \sigmab  -1)f   = (2\sigmab +1 + \rhob ) \kappa r .
\label{eqringODE3}
\end{equation}
This can be solved in general \citep[cf.][]{WiBlZa71} as we outline in appendix B. For the example above in (\ref{eqbasicstatezeta}) we have 
\begin{equation}
\sigmab = 2, \quad \rhob = - \Omega^{-2} \chi^2 (a^2 - r^2) , 
\end{equation}
inside the vortex and $\sigmab = \rhob = 0$ outside the vortex (in fact outside any vortex). We then find  
%
\begin{equation}
f =  \begin{cases}
\tfrac{5}{8} \kappa r  + \tfrac{1}{24}  \Omega^{-2} \chi^2 (r^2 - 3a^2)\kappa r, \\
\tfrac{1}{2} \kappa r \log (r/a)  
+ \tfrac{1}{8} ( 3r^2 + 2 a^2 ) \kappa r^{-1}
- \tfrac{1}{24}  \Omega^{-2}  \chi^2 a^2 (r^2 + a^2) \kappa r^{-1} . 
\end{cases}
\label{eqffarfieldmatch}
\end{equation}
Outside the vortex, solutions to the homogeneous problem (with $\sigmab$, $\rhob$ zero) are $f  = r^{-1}$ and $f = r$ and so a combination of these is chosen to make $f$ and $f_r$ continuous across $r=a$. The logarithmic term is driven by the remaining $\kappa r$ curvature term, and we need to make use of this to match to the flow in the ambient space around the vortex ring. 


\subsection{Matching and vortex ring velocity}\label{ssecringvel}

\noindent
We cannot solve for ring vortex motion based only on the local flow field calculated above, but need to appeal to how the vortex sits in the ambient space and the long-range nature of the Biot--Savart mapping from vorticity to velocity. For this it is convenient to introduce a  stream function $\psi$ for the flow $\uv$ in the vortex, excluding the axial component $u^3$. This, using (\ref{eqgovlag2}e), may be defined geometrically via 
\begin{equation}
 (u^1 \partial_r + u^2 \partial_\theta) \ip \muv = d \psiv, \quad \psiv = \psi\, dz, 
 \label{eqpsigendef0}
\end{equation}
or  in components 
\begin{equation}
u^1 =  U^1 = \mu^{-1} \partial_\theta\psi  , \quad 
u^2 = U^2 + \vb^2 = - \mu^{-1} \partial_r \psi. 
\label{eqpsigendef}
\end{equation}
Here we have
\begin{equation}
\psi =  - \int_0^r r \vb^2\, dr  - V_\cen r  \cos \theta   + \cdots, 
\end{equation}
using (\ref{eqUUpsring}a) and in line with our earlier discussion in (\ref{eqbasicstatezeta3}, \ref{eqvb2psib}). It is helpful at this point to write quantities in terms of the circulation $\Gamma$ and axial volume flux $Q$ in the vortex ring, defined by 
\begin{equation}
\Gamma = 2 \pi \Omega a^2, \quad
Q =  \tfrac{1}{4} \pi \chi a^4.
\label{eqGammaQdef} 
\end{equation}
We also repeatedly use $\kappa = \ell^{-1}$ below. Then, focusing on the outside of the vortex $ r > a $, the flow field is given by 
\begin{equation}
\psi =
 - \frac{\Gamma}{2\pi}   \log \frac{r}{a} -  V_\cen r \cos \theta + C_1 + \cdots , 
 \label{eqpsiinnermatch}
\end{equation}
from (\ref{eqbasicstatezeta}), where $C_1$ is an unimportant constant, as are the $C_j$ below.

Going now to the ambient space, we use coordinates $(R, \Theta, Z, T)$. We take the axis of the vortex ring to be given by constant $R = \ell = \kappa^{-1}$, $Z = V_{\cen} T + \cdots $ and varying $\Theta$, and we can focus on $T=0$, $Z=0$ without loss of generality. If the flow field is written as, say, 
\begin{equation}
\uv = u^R \partial_R + u^\Theta \partial_\Theta + u^Z \partial_Z,
\label{eqPsigendef0}
\end{equation}
then we may define a Stokes stream function $\Psi(R,Z)$ for the poloidal components,
\begin{equation}
(u^R \partial_R + u^Z \partial_Z) \ip \muv = d\Psiv, \quad \Psiv = \Psi \, d\Theta,
\label{eqPsigendef}
\end{equation}
corresponding to 
\begin{equation}
u^R = - R^{-1} \partial_Z \Psi, \quad
u^Z = R^{-1} \partial_R \Psi.
\end{equation}
Given the axis of the vortex is $R = \ell$, $Z=0$ in ambient coordinates, we can relate these to vortex coordinates $(\rt, \thetat, \zt)$ via
\begin{equation}
R = \ell - \rt \cos \thetat, \quad
\Theta = \kappa \zt, 
\quad Z = \rt \sin \thetat, 
\end{equation}
and a classic calculation \citep[see][]{Sa92,La94}, summarised in appendix \ref{appouterflow} for completeness, gives the outer flow for a vortex ring of circulation $\Gamma$ and radius $\ell=\kappa^{-1}$ as 
\begin{align}
\Psi 
& \simeq -  \frac{\Gamma \ell}{2 \pi} \log \rt +\frac{\Gamma \rt}{4 \pi} \left[  - \log \frac{8 \ell}{\rt}   +  1\right] \cos \thetat    + C_2 ,
\label{eqPsiouter1}
\end{align}
for points $(\rt, \thetat, \zt)$ close to the ring itself.

For matching, as $\Theta = \kappa \zt = \kappa z$ along the axis of the vortex, it turns out that we can equate $\psi$ from (\ref{eqpsigendef0}) and $\kappa \Psi$  from (\ref{eqPsigendef}) for $r>a$. Thus we take $\Psi$ from (\ref{eqPsiouter1}) and use (\ref{eqcoordtcoord}) to write $\kappa \Psi$ in terms of $(r, \theta)$ coordinates as
\begin{equation}
\kappa \Psi =-  \frac{\Gamma}{2\pi}\, \log \frac{r}{a} - \frac{\Gamma \kappa r}{4\pi} \, 
\Bigl( \,\frac{2f}{\kappa r} + \log \frac{8}{r\kappa} - 1\, \Bigr) \cos \theta + C_3 + \cdots.
\end{equation}
This should be the same as $\psi$ in (\ref{eqpsiinnermatch}) for large $r$, in the usual asymptotic matching process, which means that we need
\begin{equation}
V_\cen = \lim_{r\to\infty} \,\,\frac{\Gamma \kappa}{4\pi} \, 
\Bigl( \,\frac{2f}{\kappa r} + \log \frac{8}{r \kappa} - 1 \,\Bigr) .
\label{eqVcenlimit}
\end{equation}
Use of (\ref{eqffarfieldmatch}) and also (\ref{eqGammaQdef}) then closes the problem and fixes the velocity $V_{\cen}$,  
\begin{equation}
V_\cen =  \frac{\Gamma \kappa}{4\pi} \, 
\Bigl(  \, \log \frac{8}{a\kappa} - \frac{1}{4} - \frac{16}{3} \, \frac{Q^2}{a^2 \Gamma^2} \,\Bigr) , 
\label{eqVcenspecific}
\end{equation}
which gives the classical result of Kelvin for zero axial flow $Q=0$ and the generalisation for non-zero $Q$ as discussed in, e.g., \cite{WiBlZa71} and \cite{Sa70, Sa92}.

Note that the stream functions $\psiv$ and $\Psiv$ are nearly, but not quite, the same geometric object. To see the situation, we express, from (\ref{eqxtytztdef1}), 
\begin{equation}
\Psiv = \Psi\, d\Theta = \kappa \Psi \, d\zt 
=  \kappa \Psi \, d(z + q(r, \theta)) = \kappa \Psi\, ( dz + q_r \, dr + q_\theta\, d\theta). 
\end{equation}
The flow generated by $\Psiv$ is then given by, say $\uv_{\Psi}\ip \muv = d \Psiv$, 
and is $\uv_\Psi = u^1_\Psi \partial_z + u^2_\Psi \partial_\theta + u^3_\Psi \partial_z$ with components
\begin{align}
u^1_\Psi = \mu^{-1} \kappa\, \partial_\theta \Psi , \quad
u^2_\Psi = - \mu^{-1} \kappa \,\partial_r \Psi, \quad
u^3_\Psi = \mu^{-1} \kappa \,( \partial_r \Psi\, \partial_\theta q - \partial_\theta \Psi\, \partial_r q). 
\end{align}
The first two components correspond to the flow generated by $\psiv$ in (\ref{eqpsigendef}), but $\Psiv$ contains an extra piece over $\psiv$ since the tilting of surfaces given by $q$ means that the corresponding flow in the ambient space incorporates a part $u^3_\Psi$ of the axial flow $u^3$ in Lagrangian coordinates. In any case, outside the vortex $r\geq a$ we have that $q=0$ and so for the purposes of matching with an outer flow we can simply set $\psiv = \Psiv$ and $\psi = \kappa \Psi$, as we did above.

\subsection{The effect of axial flow}\label{ssecringeffect} 

\begin{figure}
(a)\includegraphics[scale=0.4]{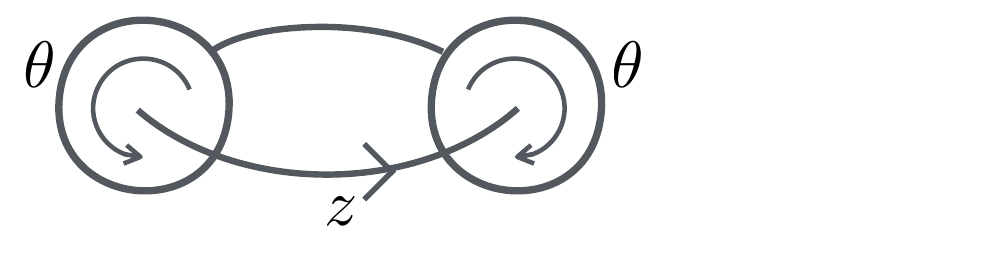}\\[5mm]
(b)\includegraphics[scale=0.4]{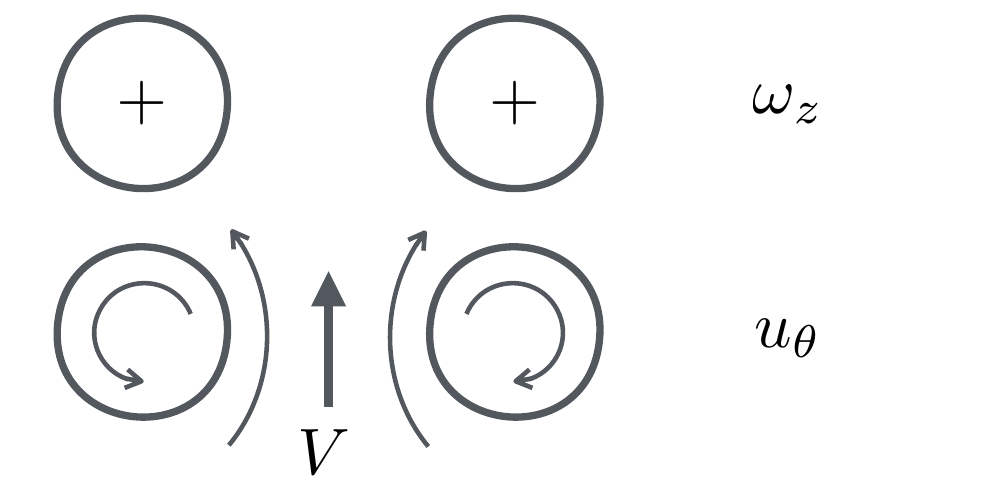}\hspace{1.0cm}
(c)\includegraphics[scale=0.4]{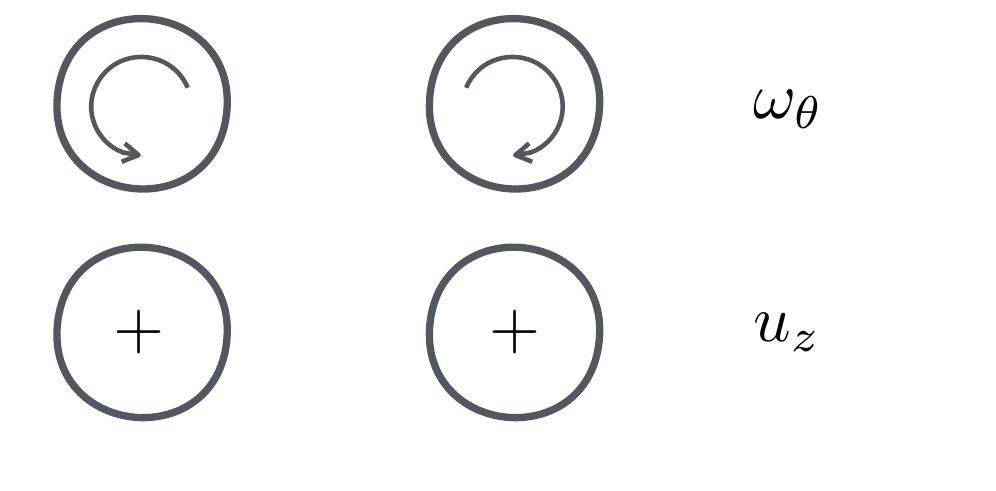}\\[5mm]
(d)\includegraphics[scale=0.4]{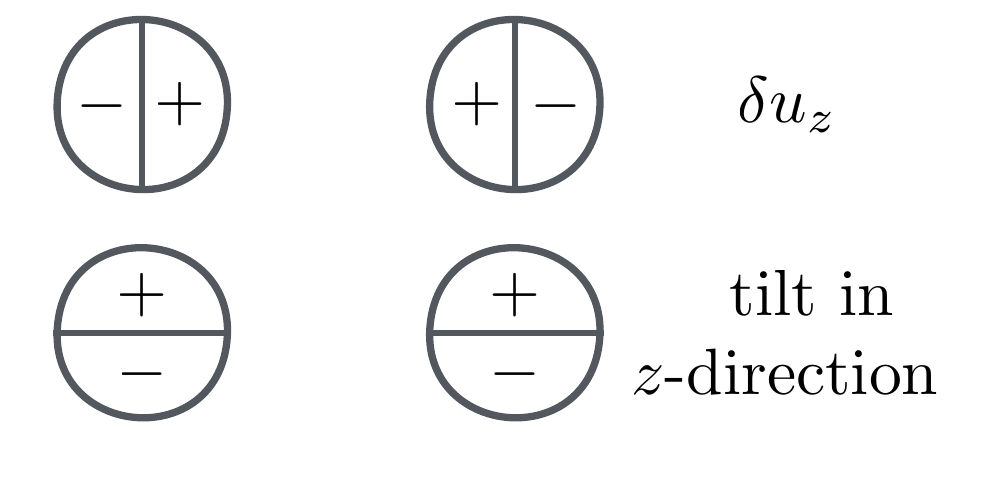}\hspace{1.0cm}
(e)\includegraphics[scale=0.4]{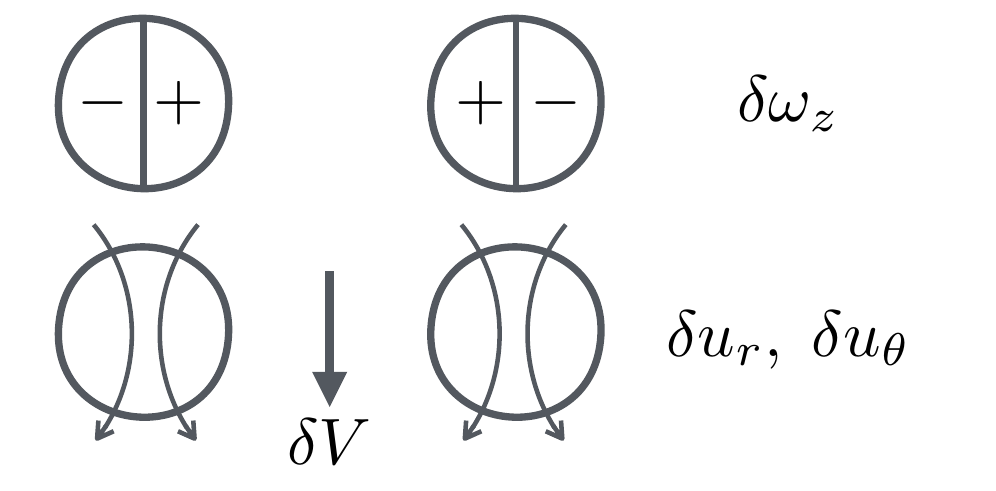}\\
\caption{Effect of axial flow on vortex ring propagation as described in the text.} 
\label{fig4}
\end{figure}

\noindent
As our appproach is based on solving the vorticity equation (rather than the Euler equation) it is of interest to give a vorticity-based explanation for the somewhat curious but well-known fact that the presence of axial flow in a vortex ring tends to reduce its forwards propagation velocity \citep{WiBlZa71,MoSa72}, resulting from the sign of the term involving  $-Q^2$ in (\ref{eqVcenspecific}). These authors discuss the effect in terms of a force balance involving the Kutta--Joukowski lift.  
If the axial flow is strong enough, in theory the direction of propagation can reverse, albeit that this is experimentally diffucult to achieve and in the presence of viscosity or other mixing processes, persistent axial motion in a vortex ring is likely to be short-lived. 

We outline the mechanism in figure \ref{fig4}, using letter subscripts for components to aid intuition. Panel (a) is to remind us of the $(\theta, z)$ elements of the coordinate system. Panel (b) shows the axial vorticity $\omega_z$ and corresponding flow $u_\theta$ for a ring without axial flow, propagating with velocity $V$, say. In panel (c) we introduce an axial flow, corresponding to distributions of $u_z$ and $\omega_\theta$. Now, because of conservation of angular momentum about the major axis of the vortex ring (the $Z$-axis), the axial flow $u_z$ is stronger on the inner side of the vortex core and weaker on the outer side. This is depicted in panel (d) with $\delta u_z$ showing $u_z$ minus its mean value, positive on the inner side, negative on the outer side. The effect of this is to tilt the vorticity surfaces $\Sc(z,t)$ (see figure \ref{fig3}) in the $z$-direction, as captured by the function $q$ in our analysis; see (\ref{eqcoordtcoord}, \ref{eqringqrel}). The tilting \citep[remarked on in][]{WiBlZa71} of the curves of $\omega_\theta$ in panel (c) results in additional axial vorticity $\delta \omega_z$ as depicted in panel (e). This vorticity has a dipolar distribution within the ring, creating local velocity components $\delta u_r$, $\delta u_\theta$ and giving an additional velocity $\delta V$ to the whole ring as depicted, which tends to oppose the original velocity $V$ in panel (b).


\section{General development}\label{secfullgeneral}

\noindent
We now leave aside vortex rings and return to the broader development, to see how far we can take this geometric framework for  vortex dynamics, allowing the most general scalings  without specifying a particular application. We pick up the discussion from the end of \S\ref{seclagcoord} and find it clearest to proceed in two steps. The first step, in the present section, is to develop equations without the use of any assumption of slenderness. In the second step, in \S\ref{secgeneralslender}, we apply the limit of a slender vortex, set out in \S\ref{secslender}, for further simplifications.


\subsection{Fluid system in general}

\noindent
Our focus now is the governing equations (\ref{eqgovlag1}), in which we have applied the geometric conditions (\ref{eqzetalag1}) and (\ref{eqvlag1}). We also have the centreline equations in (\ref{eqcurvebasisdt1}) and the definition of the background flow  in \S\ref{sseclagcoordback}, but no slenderness assumption; everything is exact. We will take the mean or fluctuating component, with respect to $\theta$, for each equation in (\ref{eqgovlag1}). For coherence, we gather together the final results in one place here, with more explanation following, namely,
\begin{subequations}\label{eqgovlag3}
 \begin{align}
 \partial_{t}  \xib_2 & + \partial_{z} (\vb^3 \xib_2 )  =  \xib_3\,  \partial_{z} \vb^2 , \\
 \xib_3(r) & = \partial_r \nub_2(r), \\
\xib_2 & = \partial_z \nub_1 - \partial_r \nub_3, 
\\
 \nub_i - \Upsilonb_i & = \gb_{i2} \vb^2 + \gb_{i3} \vb^3, 
\\
- \Xi'_1
& = (\partial_\theta g'_{32} - \partial_z g'_{22}) \vb^2 
+(\partial_\theta  g'_{33} - \partial_z g'_{23}) \vb^3  
- g'_{22} \partial_z \vb^2 - g'_{23} \partial_z \vb^3, 
\\
- \Xi'_2 
& = \partial_z (g'_{12} \vb^2 + g'_{13}\vb^3 ) 
- \partial_r ( g'_{32} \vb^2 +  g'_{33} \vb^3  ) , \\
 - \Xi'_3
& = (\partial_r g'_{22} - \partial_\theta g'_{12}) \vb^2 
+(\partial_r g'_{23} - \partial_\theta g'_{13}) \vb^3  
+ g'_{22} \partial_r \vb^2 + g'_{23} \partial_r \vb^3, 
\\
 \partial_t \mub & +  \partial_z (\mub \vb^3)   = 0 , \\
\partial_t \mu'  & + \vb^2 \partial_\theta \mu'  + \partial_z (\mu' \vb^3)  = 0 .
  \end{align}
\end{subequations}
We have simply repeated equation  (\ref{eqgovlag1}a) to give (\ref{eqgovlag3}a). 
We now turn to the equations (\ref{eqgovlag1}b--d) linking the vorticity $\xiv$ to the momentum $\nuv$.
Taking the mean of (\ref{eqgovlag1}b, d) shows that (\ref{eqgovlag3}b) holds. 
Thus given that the basic vorticity distribution $\xib_3(r)$ is an input to our problem, a function we can specify arbitrarily in (\ref{eqzetalag1}c) at the outset, the corresponding mean momentum $\nub_2(r)$ is fixed. 
Next, the mean of (\ref{eqgovlag1}c) results in (\ref{eqgovlag3}c). 

We now take the fluctuating components of (\ref{eqgovlag1}b--d) and observe that,  as $\xiv = \xivb$, the resulting left-hand sides give zero and we have 
\begin{equation}
0 = \partial_\theta \nu'_3 - \partial_z \nu'_2, \quad
0 = \partial_z \nu'_1 - \partial_r \nu'_3, \quad
0 = \partial_r \nu'_2 - \partial_\theta \nu'_1 
\tor d\nuv' = 0 .
\label{eqPhisolve}
\end{equation}
We could use this to introduce a potential, $\nuv' = d\varphi'$, 
but this does not seem to be helpful. Instead we will use (\ref{eqPhisolve}) for another purpose, shortly. 

In what follows we note that the geometric conditions (\ref{eqvlag1}) are expressed in terms of $\vv$; however the equations we deal with next involve $\uv$ through the related quantities $\nuv$ and $\xiv$. To make best use of  (\ref{eqvlag1}) we need to replace all quantities related to $\uv$ with those related to $\vv$ by applying (\ref{eqschema}, \ref{eqschemarels}).  In other words we will subtract the background flow fields $\Uv$, $\Upsilonv$, $\Xiv$ from the fluid flow fields $\uv$, $\nuv$, $\xiv$, to work in terms of the relative flow fields $\vv$, $\piv$, $\varpiv$.


We first turn to (\ref{eqgovlag1}e) which links momentum $\nuv$ to the full flow field $\uv$, that is $\nuv = \uv_\flat$. 
We subtract the background momentum $\Upsilonv = \Uv_\flat $ from each side, and find 
\begin{equation}
\nuv - \Upsilonv = \piv = 
\vv_\flat = \vvb_\flat
\tor
\nu_i - \Upsilon_i =  g_{ij} \vb^j , 
\label{eqnuUpsilonlink1}
\end{equation}
whose mean components give (\ref{eqgovlag3}d)  for $i=1$, $2$, $3$. 
Taking instead the fluctuating components of (\ref{eqnuUpsilonlink1}) gives 
\begin{equation}
\nuv' - \Upsilonv' = \piv' =  (\vv_\flat)' = (\vvb_\flat)' = \gv'(\vvb, \cdot) 
\tor
\nu'_i - \Upsilon'_i =   g'_{ij} \vb^j.
\label{eqnuUpsilonlink3}
\end{equation}
It does not appear too helpful to track the $\nu'_i$ in our calculations and so we now apply the exterior derivative $d$ to these equations and use $d\nuv'=0$ (\ref{eqPhisolve}) to eliminate $\nuv'$. We can also replace $d\Upsilonv' = \Xiv'$, which brings in the background flow vorticity, to give
\begin{equation}
- \Xiv' = \varpiv' =  d \piv'  = d [(\vvb_\flat)'] = d[ \gv'(\vvb, \cdot)]  .
\end{equation}
Writing this out in components yields the three equations (\ref{eqgovlag3}e--g). 
Note that these equations are not independent: since $\Xiv' = d \Upsilonv'$, we have $d\Xiv' = 0$, or
\begin{equation}
\partial_r \Xi'_1 + \partial_\theta \Xi'_2 + \partial_z \Xi'_3 = 0 . 
\label{eqdXiprime0}
\end{equation}
We could therefore discard one of  (\ref{eqgovlag3}e--g) but it is not clear which, so we retain all three. Note that (\ref{eqgovlag3}e--g) have the physical interpretation that there is no fluctuating vorticity $\xiv'$ in the fluid flow, as per (\ref{eqvlag1}), so that any fluctuating vorticity $\Xiv'$ in the background flow must be accounted for on the right-hand side. 

Finally we turn to the incompressibility condition (\ref{eqgovlag1}f); again this refers to the full flow $\uv$ and not the relative flow $\vv$. However we use $\divv \uv = \divv \Uv + \divv \vv = 0 $, which together with (\ref{eqmuidentity}), gives 
\begin{equation}
\vb^2\partial_\theta \mu   + \partial_z (\mu \vb^3) + \partial_t \mu = 0 . 
\end{equation}
Taking the mean and fluctuating parts gives (\ref{eqgovlag3}h, i).

This completes what we can achieve in the general Lagrangian system without the slenderness assumption. We have traded references to the flow $\uv$ with the relative flow $\vv$ and so brought in the background flow $\Uv$, through its momentum $\Upsilonv$, vorticity $\Xiv$ and divergence. All the equations are exact and the system (\ref{eqgovlag3}) discussed in this section, plus the centre line equations in (\ref{eqcurvebasisdt1}) and the background quantities discussed in \S\ref{sseclagcoordback} (we spared ourselves the effort of expanding these fully but this is possible in principle), provide a complete and exact set of local governing equations in the Lagrangian framework.


\section{General development in the slender limit}\label{secgeneralslender}

\noindent
In this section we apply the limit of a slender vortex and the scalings detailed in \S\ref{ssecslenderdef} to the full system, of the centre line equations (\ref{eqcurvebasisdt1}), the background flow in \S\ref{sseclagcoordback} and the fluid equations (\ref{eqgovlag3}). In fact the centre line equations (\ref{eqcurvebasisdt1}) are easily dealt with as they do not simplify; every term is of the same size, namely $O(\eps)$. We also recall our use of the shorthand notation ${}+ \cdots$ to mean ${}+O(\eps^2)$.

\subsection{Background flow in slender limit}\label{secbackgroundslender}

\noindent
We continue the discussion from \S\ref{ssecslendermetric}, where we had general forms for $\Qv$, $\gv$ and $\muv$ in the slender vortex limit. We make use of (\ref{eqcoordgenq}) to calculate quantities pertaining to the background flow. First, applying $(\Qv^{-1})^T$ to the components $(U,V,W)$  of $\Uv$ in (\ref{eqfullUgen0b}) gives the corresponding contravariant components:
\begin{subequations}\label{eqUbasic2}
\begin{align}
\alpha U^1 & =   U_\cen \yt_\theta   - V_\cen \xt_\theta  - (S_\cen - \tau W_\cen) \rt \rt_\theta + J^{-1} W_\cen (\xt_\theta \yt_z - \xt_z \yt_\theta) + \xt_t \yt_\theta - \xt_\theta \yt_t  + \cdots ,  
\label{eqU1basic2}\\
\alpha U^2 & = - U_\cen \yt_r  + V_\cen \xt_r + (S_\cen - \tau W_\cen) \rt \rt_r + J^{-1} W_\cen (\xt_z \yt_r - \xt_r \yt_z) + \xt_r \yt_t - \xt_t \yt_r + \cdots , 
\label{eqU2basic2}\\
\alpha U^3 & = J^{-1} \alpha W_\cen 
 - J^{-1}\alpha  (U_{\cen s} - \tau V_\cen) \xt  - J^{-1}\alpha  (V_{\cen s}   + \tau U_\cen) \yt 
  + U_\cen (\yt_r q_\theta - \yt_\theta q_r ) - V_\cen (\xt_r q_\theta - \xt_\theta q_r) + \cdots  . 
\label{eqU3basic2}
\end{align}
\end{subequations}
Here the expression $S_\cen -  h^{-1} \tau W$ in (\ref{eqUbasic1}a, b) is replaced by $S_\cen - \tau W_\cen + \cdots$. 
Although it is useful to exhibit these components here in (\ref{eqUbasic2}), they are not needed much in the forthcoming development; the momentum $\Upsilonv = \Uv_\flat$ is much more relevant and is given by, using (\ref{eqcoordgenq}), 
\begin{subequations}\label{eqUpsilonbasic1}
\begin{align}
\Upsilon_1  &   = U_\cen \xt_r   + V_\cen \yt_r  - S_\cen (\xt_r \yt  - \xt \yt_r) + J W_\cen q_r + \xt_r \xt_t + \yt_r \yt_t +  \cdots ,   \\
\Upsilon_2 &  = U_\cen \xt_\theta   + V_\cen \yt_\theta  - S_\cen (\xt_\theta \yt  - \xt \yt_\theta ) + J W_\cen q_\theta + \xt_\theta \xt_t + \yt_\theta \yt_t +  \cdots ,   \\
\Upsilon_3  &  = J h^2 W_\cen  -   J(U_{\cen s} -  2 \tau V_\cen) \xt   - J( V_{\cen s} +  2 \tau U_\cen)  \yt + U_\cen \xt_z + V_\cen \yt_z  +  \cdots.
\end{align}
\end{subequations}
The background vorticity, $\Xiv = d \Upsilonv$, is 
\begin{subequations}\label{eqZetabasic2}
\begin{align}
\Xi_1  &   
= - 2J   (U_{\cen s} - \tau V_\cen)   \xt_\theta -   2 J (V_{\cen s}+ \tau U_\cen) \yt_\theta   - 2 J \kappa W_\cen \xt_\theta + \cdots , 
  \\
\Xi_2  &  
=   2J   (U_{\cen s} - \tau V_\cen)   \xt_r + 2 J (V_{\cen s}+ \tau U_\cen) \yt_r    + 2 J  \kappa W_\cen \xt_r + \cdots, \\
\Xi_3 &  
=  2 \alpha S_\cen   +  \xt_\theta \xt_{rt} - \xt_r \xt_{\theta t}  + \yt _\theta \yt_{r t}-  \yt _r \yt_{\theta t}    + \cdots.
\end{align}
\end{subequations}
The divergence of the background flow is given by 
%
\begin{equation}
\divv \Uv = \mu^{-1} \bigl[ (\mu U^1)_{r} +   (\mu U^2)_{\theta}  +  (\mu U^3)_{z} \bigr]  .
\end{equation}
The background flow is that we had earlier in (\ref{eqUbasic1}) for the vorticity coordinate system, with the addition of the terms in $\Uv_{\flex}$ in (\ref{eqUflexdef}). The divergence of the present background flow $\Uv$ is then given by the divergence calculated earlier in (\ref{eqdivvUv1a}), as it is a scalar independent of coordinates, plus the divergence of $\Uv_{\flex}$, which is $\divv \Uv_{\flex} = \alpha^{-1} \alpha_t + \cdots $. Putting these together confirms the exact result that 
\begin{equation}
\divv \Uv =  \mu^{-1} \mu_t = J^{-1} J_t + \alpha^{-1} \alpha_t +  h^{-1} h_t = J^{-1} J_t + \alpha^{-1} \alpha_t + \cdots , 
\label{eqdivvUv2} 
\end{equation}
from (\ref{eqmuidentity}). Thus the background flow, as we have defined it, generally carries an $O(\eps)$ divergence, $O(1)$ momentum and $O(\eps)$ vorticity. 
 

\subsection{Fluid system in slender limit}

\noindent 
We now apply the slender limit in \S\ref{secslender} to the fluid equations in (\ref{eqgovlag3}) and make use of results in \S\ref{secbackgroundslender}. We expand quantities in the slenderness parameter $\eps$ to leading and first order, with some exceptions. We use a relaxed Roman index, $\oo$, $\ii$, $\ii\ii$, $\ii\ii\ii$, $\ii\upsilon$, $\ldots$, to indicate the order in $\eps$, though  fortunately we only need the first two of these. We will expand numerous quantities in this way, giving the $O(1)$ term and then the $O(\eps)$ term, for example,
\begin{equation}
\nub_1 = \nub_{1\oo} + \nub_{1\ii} + \cdots, \quad
\Xi_1' = \Xi'_{1\oo} + \Xi'_{1\ii} + \cdots, \quad
g_{12} = g_{12 \oo} + g_{12 \ii} + \cdots .
\end{equation}
Quantities we will not need to expand are the centre line variables,
\begin{equation}
J,\, \kappa, \, \tau,\, U_\cen,\, V_\cen,\, W_\cen, \,S_\cen,  
\end{equation}
and the axial vorticity distribution $\xi_3 = \xib_3(r)$, which is an input to the problem, together with the corresponding mean momentum $\nub_2(r)$ related by (\ref{eqgovlag3}b). In other words we take 
\begin{equation}
\xib_3(r) = \xib_{3\oo}(r), \quad
\nub_2(r) = \nub_{2\oo}(r) ,
\label{eqxib3oo}
\end{equation}
with no first order contributions.

We first look at the background flow from \S\ref{secbackgroundslender} in the slender limit and note the following simplifications from (\ref{eqPhisolve}, \ref{eqUpsilonbasic1}, \ref{eqZetabasic2}),
\begin{subequations}\label{eqslendsimps1}
\begin{align}
\nu_{3\oo} & = \nub_{3\oo}, \quad \nu'_{3\oo} = 0 , \\
\Upsilon_{2\oo} & = \Upsilon'_{2\oo} ,   \quad \Upsilonb_{2\oo}= 0 , \\
\Upsilon_{3\oo} & = \Upsilonb_{3\oo}(z,t) = JW_\cen, \quad \Upsilon'_{3\oo}=0, \\
\Xi_{1\oo} & = \Xi_{2\oo}  = \Xi_{3\oo} = 0 , \\
\Xi_{1\ii} & = \Xi'_{1\ii}, \quad \Xib_{1\ii} = 0 .
\end{align}
\end{subequations}
Next we consider the metric components $g_{ij}$ from (\ref{eqmetricslender}), which satisfy 
\begin{subequations}\label{eqmetricsimps1}
\begin{align}
g_{13\oo}  & = g_{31\oo} = 0 ,  \\
g_{23\oo}  & = g_{32\oo} = 0 ,  \\
g_{33\oo} & = \gb_{33\oo}(z,t) = J^2, \quad g'_{33\oo} = 0 , 
\end{align}
\end{subequations}
and we have $\alpha$ and $\mu$ from (\ref{eqmuslender}--\ref{eqalphaslender1}), which do not simplify. 


\subsection{Leading order fluid system in slender limit}

\noindent
We now follow the framework of our discussion of the vortex ring in section \ref{secring}. At leading order a vortex filament is not curved and we need the analogue of the straight column in \S\ref{ssecringstraight}. This is given by setting $\eps$, $\kappa$ and all quantities labelled $\ii$ to zero, in other words picking up only the non-trivial equations in the system (\ref{eqgovlag3}) taken at $O(1)$, which are
\begin{subequations}\label{eqgovlag4}
 \begin{align}
\xib_{3\oo}(r) & = \partial_r \nub_{2\oo}(r), \\
\xib_{2\oo} & =   - \partial_r \nub_{3\oo}, 
\\
\nub_{1\oo} - \Upsilonb_{1\oo} & = \gb_{12\oo} \vb^2_{\oo}   , 
\\
\nub_{2\oo} & = \gb_{22\oo} \vb^2_{\oo}   , 
\\
\nub_{3\oo} - \Upsilonb_{3\oo} & =  \gb_{33\oo} \vb^3_{\oo} , 
\\
 0
& = (\partial_r g'_{22\oo} - \partial_\theta g'_{12\oo}) \vb^2_{\oo}+ g'_{22\oo} \partial_r \vb^2_{\oo} ,
\\
 \alpha'_{\oo}  & = 0 . 
\end{align}
\end{subequations}
We work from the system (\ref{eqgovlag3}), making use of the information in (\ref{eqslendsimps1}, \ref{eqmetricsimps1}). 
First, equation (\ref{eqgovlag3}a) has no zero order terms, while (\ref{eqgovlag3}b, c) result in 
(\ref{eqgovlag4}a, b), the first of which fixes $\nub_{2\oo}(r)$ in terms of the input axial vorticity $\xib_{3\oo}(r)$, for all time; see (\ref{eqxib3oo}). On the other hand the axial flow $\nub_{3\oo}(r)$ and corresponding azimuthal vorticity $\xib_{2\oo}(r)$ will generally evolve on the long space and time scale (see (\ref{eqgovlag5}a) below) but are related by (\ref{eqgovlag4}b).

Equation (\ref{eqgovlag3}d) links momenta and velocities, and results in (\ref{eqgovlag4}c--e); curiously, the first of these is purely diagnostic and gives $\nub_{1\oo}$, which appears nowhere else in our analysis and may be dropped. The remaining two link the azimuthal and axial mean momenta to the corresponding velocity fields, and are analogous to (\ref{eqnu23relate}) for our vortex ring calculation. Note (\ref{eqgovlag4}e) is 
\begin{equation}
\nub_{3\oo} = J (W_\cen + J \vb^3_{\oo}), 
\end{equation}
as might be expected for the leading order axial momentum. Of equations (\ref{eqgovlag3}e--g) for the fluctuating vorticity, only the third is non-trivial at leading order, and results in (\ref{eqgovlag4}f), which can also be written  as 
\begin{equation}
\frac{1} {g'_{22\oo} }
\, \bigl( 
\partial_r g'_{22\oo} - \partial_\theta g'_{12\oo} \bigr)  
= - \frac{\partial_r \vb^2_{\oo} }{  \vb^2_{\oo}}\,  .
\label{eqawkflucexp1} 
\end{equation}
The fluctuating part of the left-hand side must be zero, which can be viewed as a constraint on the geometry and as a solvability condition for the flow $\vb^2_{\oo}$.

 From  (\ref{eqgovlag3}h, i) we are left  only with  $\mu'_{\oo}=0$ and from (\ref{eqmuslender}--\ref{eqalphaslender1}) this implies that (\ref{eqgovlag4}g) holds, where we recall that $\alpha$ gives the area element in $(r,\theta)$ surfaces $\Sc(z,t)$. The condition (\ref{eqgovlag4}g) can also be written, awkwardly, in terms of metric elements, 
\begin{equation}
\alpha'_{\oo}  =  \bigl( \sqrt{g_{11\oo} g_{22\oo} - g_{12\oo} g_{12\oo}} \bigr)^{\prime} = 0 .
\end{equation}

The system (\ref{eqgovlag4}) represents inviscid fluid flow in the $(r,\theta)$ plane, with also perpendicular flow, involving only planar metric elements $g_{11\oo}$,  $g_{12\oo}$ and $g_{22\oo}$ (and the trivial $\gb_{33\oo}=J$). The system does not include the curvature $\kappa$ and does not involve $(z,t)$ except as parameters. Thus a solution will generally vary along the vortex, evolving with respect to the slow $(z,t)$ coordinates by the next order system we discuss in the next section. While (\ref{eqgovlag4}a--e) essentially just define quantities in terms of other quantities, it is satisfying equations (\ref{eqgovlag4}f, g) that correspond to imposing the incompressible Euler equations in the $(r, \theta)$ plane. These are easily satisfied for the case of axisymmetric flow (for which $g'_{ij\oo}=0$, $\alpha'_{\oo}=0$) as in \S\ref{ssecringstraight}, but unwieldy in general. Nonetheless, from the perspective of the theoretical development, this set up is extremely flexible and allows any two-dimensional flow with the topology of cylindrical vortex surfaces; for example, this could include one of a pair of vortices interacting closely, with distorted cores, or one lobe of a vortex dipole. In short, for a vortex filament immersed in a general irrotational external flow, we allow arbitrary core distortions, steady on the turnover time-scale but varying arbitrarily in the slow variables $(z,t)$, subject only to the slenderness assumption. 

In our development we have discarded equations for the fluctuating momentum $\nu'_i$, in favour of equations for the fluctuating vorticity $\Xi'_i$, for example (\ref{eqgovlag3}e--g) and here (\ref{eqgovlag4}f). However it is useful to set these out here as
\begin{equation}
\partial_r \nu'_{2\oo}  - \partial_\theta \nu'_{1\oo} =0 , 
\end{equation}
from (\ref{eqPhisolve}), and
\begin{subequations}
\begin{align}
\nu'_{1\oo} - \Upsilon'_{1\oo}  & = g'_{12\oo} \vb^2_{\oo} ,  \\
\nu'_{2\oo} - \Upsilon'_{2\oo}  & = g'_{22\oo} \vb^2_{\oo} ,
\end{align}
\end{subequations}
from (\ref{eqnuUpsilonlink3}). These, together with $\Xi_{3\oo}= 0 $ in (\ref{eqslendsimps1}), led to (\ref{eqgovlag4}f).



\subsection{First order fluid system in slender limit}

\noindent
We now look at the governing equations (\ref{eqgovlag3}) at the level of $O(\eps)$ terms in the slender vortex limit. Our goal is to find all the equations that involve the leading order `$\oo$' fields that feature in (\ref{eqgovlag4}) above, particularly $\vb^2_{\oo}$ and $\vb^3_{\oo}$. This will provide equations that govern the evolution of these fields on the slow $(z,t)$ scales and that also constrain and define the vortex geometry, as seen in the case of the vortex ring calculation earlier, where the order $\eps$ terms led to the two differential equations giving the distortions arising from curvature and ultimately the velocity of the ring after matching to the external flow. What we do not want to do is to introduce first order fields, that is, the `$\ii$' fields. We wish to manage all the flow and vorticity quantities at leading order, and only allow the geometry to have order $\eps$ distortions and `$\ii$' terms, that is in the metric entries $g_{ij}$ and in $\mu$. 
Bearing these points in mind, the relevant equations are
\begin{subequations}\label{eqgovlag5}
 \begin{align}
 \partial_{t}  \xib_{2\oo} & + \partial_{z} (\vb^3_{\oo}  \xib_{2\oo}  )  =  \xib_{3\oo} \,  \partial_{z} \vb^2_{\oo}  , 
 \\
- \Xi'_{1\ii}
& = (\partial_\theta g'_{32\ii} - \partial_z g'_{22\oo}) \vb^{2}_{\oo} 
+(\partial_\theta  g'_{33\ii}) \vb^3_{\oo}
- g'_{22\oo} \partial_z \vb^2_{\oo} , 
\\
- \Xi'_{2\ii} 
& = \partial_z (g'_{12\oo} \vb^2_{\oo} ) 
- \partial_r ( g'_{32\ii} \vb^2_{\oo} +  g'_{33\ii} \vb^3_{\oo}  ) , 
\\
 - \Xi'_{3\ii}
& = (\partial_r g'_{22\oo} - \partial_\theta g'_{12\oo}) \vb^2_{\ii}+ g'_{22\oo} \partial_r \vb^2_{\ii} \notag
\\
& + (\partial_r g'_{22\ii} - \partial_\theta g'_{12\ii}) \vb^2_{\oo}  + g'_{22\ii} \partial_r \vb^2_{\oo} 
+(\partial_r g'_{23\ii} - \partial_\theta g'_{13\ii}) \vb^3_{\oo}   
 + g'_{23\ii} \partial_r \vb^3_{\oo} , 
\\
\partial_t \mub_{\oo}   & +  \partial_z (\mub_{\oo} \vb^3_{\oo})  = 0 , \\
\mu'_{\ii} & = 0 .
\end{align}
\end{subequations}
We look through the equations in (\ref{eqgovlag3}) at order $\eps$, in turn, noting (\ref{eqslendsimps1}, \ref{eqmetricsimps1}). 
Equation (\ref{eqgovlag3}a) is unchanged to become (\ref{eqgovlag5}a), while (\ref{eqgovlag3}b) gave (\ref{eqgovlag4}b) and nothing further, from (\ref{eqxib3oo}). Eqations (\ref{eqgovlag3}c, d) at $O(\eps)$  involve the higher order `$\ii$' fields and so we do not write these down. We found that (\ref{eqgovlag3}e, f) had no terms at $O(1)$ and so did not feed into (\ref{eqgovlag4}), but these now give something non-trivial at $O(\eps)$, namely (\ref{eqgovlag5}b, c). These two equations are related by 
\begin{equation}
\partial_r \Xi'_{1\ii} + \partial_\theta \Xi'_{2\ii} = 0 , 
\label{eqdXiprime1}
\end{equation}
from (\ref{eqdXiprime0}). For (\ref{eqgovlag3}g), we obtained a useful equation (\ref{eqgovlag4}f) at $O(1)$ and the $O(\eps)$ terms give (\ref{eqgovlag5}d), which at first sight does not appear relevant as it involves the higher order flow field $\vb^2_{\ii}$. However making use of (\ref{eqgovlag4}f) means that  (\ref{eqgovlag5}d) can be written as 
\begin{align}
  \frac{1}{g'_{22\oo}} \, 
  \bigl[  \Xi'_{3\ii}+ (\partial_r g'_{22\ii} - \partial_\theta g'_{12\ii}) \vb^2_{\oo}  
+(\partial_r g'_{23\ii} - \partial_\theta g'_{13\ii}) \vb^3_{\oo}   
+ g'_{22\ii} \partial_r \vb^2_{\oo}  + g'_{23\ii} \partial_r \vb^3_{\oo} 
\bigr] 
& = \frac{1}{\vb^2_{\oo}} \, \bigl( \vb^2_{\ii} \, \partial_r \vb^2_{\oo} - \vb^2_{\oo} \, \partial_r \vb^2_{\ii} \bigr)
\label{eqgovlag5p}
\end{align}
(cf.\ the leading order (\ref{eqawkflucexp1})). 
So, the fluctuating part of the left-hand side must be zero,    
 \begin{align}
   \Bigl (\, \frac{1}{g'_{22\oo}} \, 
  \bigl[  \Xi'_{3\ii}+ (\partial_r g'_{22\ii} - \partial_\theta g'_{12\ii}) \vb^2_{\oo}  
+(\partial_r g'_{23\ii} - \partial_\theta g'_{13\ii}) \vb^3_{\oo}   
+ g'_{22\ii} \partial_r \vb^2_{\oo}  + g'_{23\ii} \partial_r \vb^3_{\oo} 
\bigr] \,\Bigr)' 
& = 0, 
\label{eqawkflucexp2} 
\end{align}
and this constrains the geometry, as a sovability condition for the $\vb^2_{\ii}$ field. Note that if $g'_{22\oo}=0$ we need to reconsider (\ref{eqgovlag5}d).

Finally (\ref{eqgovlag3}h) is unchanged as (\ref{eqgovlag5}e) while (\ref{eqgovlag3}i) yields (\ref{eqgovlag5}f) as $\mu'_{\oo} = J \alpha'_{\oo}=0$ (see (\ref{eqgovlag4}g)). Since $\mu = J h \alpha + \cdots$,  we have $\mu_{\ii} = J (\alpha_{\ii} - \kappa \alpha_{\oo} \xt_{\oo})$ and taking the fluctuating part gives
\begin{equation}
\alpha'_{\ii} = \kappa \alphab_{\oo} \xt'_{\oo} .
\end{equation}
Since $\xt'_{\oo}$ generally has a fluctuating part (as particles circulate around the vortex core), what this equation indicates is that any curvature $\kappa\neq0$ drives a fluctuating part of the area element in the surfaces $\Sc(z, t)$ in figure \ref{fig3}, so preserving volume conservation in the ambient space. 

We have completed our discussion of the governing equations (\ref{eqgovlag5}) at order $\eps$. In addition, it is worth noting some of the fluctuating momentum equations at this order, namely 
\begin{subequations}\label{eqflucmomiii}
\begin{align} 
 \partial_\theta \nu'_{3\ii} -  \partial_z \nu'_{2\oo} & = 0 , \\
 \partial_z \nu'_{1\oo}  - \partial_r \nu'_{3\ii} & = 0, 
\end{align}
\end{subequations}
from the first two of (\ref{eqPhisolve}), and
\begin{align} 
\nu'_{3\ii} - \Upsilon'_{3\ii} & = g'_{32\ii} \vb^2_{\oo} +g'_{33\ii} \vb^3_{\oo}, 
\label{eqflucmomiv}
\end{align}
from the last of (\ref{eqnuUpsilonlink3}). 
  

\subsection{Vortex ring motion and area waves revisited}

\noindent
In this section we revisit how our solution for vortex rings in section 6 maps onto the above general framework. The vortex ring simplifications in (\ref{eqcurved2}) make many quantities identically zero, in particular, 
\begin{equation}
\Xiv=0, \quad \Upsilon_3=0, \quad \nu'_3 =0 , 
\label{eqringsimps}
\end{equation}
together with no torsion $\tau=0$ and the symmetries given by $\partial_t$, $\partial_z = 0$. In other words, the background flow is uniform, carrying no vorticity (\ref{eqZetabasic2}) and no axial momentum (\ref{eqUpsilonbasic1}c), and the fluctuating axial momentum is zero from (\ref{eqPhisolve}). It is easy to confirm that the leading order equations (\ref{eqgovlag4}) are satisfied from the vortex column solution in \S\ref{ssecringstraight}.
 
Then, we can identify three key equations from the development in section \ref{ssecringstruct}. The first is the identification of the tilt $q$ in (\ref{eqringqrel}), which comes from the axial fluctuating momentum in (\ref{eqringnurel}c). This corresponds to applying (\ref{eqflucmomiv}), which is easy to do in view of (\ref{eqringsimps}), and making use of $g'_{32\ii}$ and $g'_{33\ii}$. In a more general situation this would correspond to use of  (\ref{eqgovlag5}b,~c) for the first two fluctuating vorticity components; for a vortex ring these integrate to give (\ref{eqflucmomiv}). The second key equation is the first of our two ODEs, (\ref{eqringODE1}), and this arises from the axial vorticity equation (\ref{eqgovlag5}d). Since $g'_{12\oo}$, $g'_{22\oo} = 0$, the first order field $ \vb^2_{\ii}$ is not involved, and so we do not need to reach for the daunting (\ref{eqawkflucexp2}). The use of (\ref{eqgovlag5}d) involves $g'_{12\ii}$, $g'_{22\ii}$, and $g'_{13\ii}$,  $g'_{23\ii}$ to include coupling to $q$. The final key equation is the second of our two ODEs, (\ref{eqringODE2}), which straightforwardly corresponds to (\ref{eqgovlag5}f), that the coordinate system have an action--angle form at order $\eps$. 

Thus (\ref{eqgovlag5}b--d, f) are all involved in the problem of vortex ring motion, and in particular all the first order fluctuating metric terms $g'_{ij\ii}$ present in (\ref{eqgovlag5}) are needed for the calculations. Equations (\ref{eqgovlag5}a, e) are trivially satisfied in the vortex ring problem since $\partial_z$, $\partial_t=0$, and so these do not play a role. 

However there is the other straightforward application of this general system, to area waves on an axisymmetric, columnar vortex, studied by \cite{ChGi21} in the present framework and building on  earlier studies \citep{LuAs89, Le94, Ma02}. In this case $\partial_\theta=0$ so all fluctuating quantities are zero, but there is a wave-like dependence on $(z,t)$ coordinates. We omit the details, but equation (\ref{eqgovlag5}a) is crucial in generating azimuthal vorticity $\xib_{2\oo}$ through variations in angular velocity $\vb^2_{\oo}$, and so axial flow $\vb^3_{\oo}$. Meanwhile (\ref{eqgovlag5}f) imposes volume conservation in the vortex core as bulges and indentations, essentially axisymmetric Kelvin waves, propagate. 

These considerations indicate that the framework in this paper, while complicated, is in a sense minimal for capturing both vortex ring dynamics (and hence thin filament dynamics) and area wave dynamics, on vortices of arbitrary cross section with order one velocities and non-trivial core distortions. Specifically, all the equations in (\ref{eqgovlag4}, \ref{eqgovlag5}) are needed, and all the metric terms $g'_{ij\ii}$ that are present above play a role. 



\section{Discussion}\label{secdiscuss}

\noindent
In this paper we have taken a novel approach to a general class of problems involving slender vortices in inviscid, incompressible fluid dynamics in three dimensions. We have developed a coordinate system where the surfaces on which vortex lines spiral are given by a coordinate $r = \mathrm{const.}$ and the flow and vorticity have an action--angle form relative to a coordinate $\theta$. A third coordinate $z$ increases along the vortex, which is localised near to an evolving curve $\Cc(t)$. We applied this framework to the motion of a slender vortex ring, and together with matching using the Biot--Savart integral for the flow outside the ring, regained classic results for ring motion with and without axial flow \citep{Sa70,Sa92,WiBlZa71}. 

In our geometrical set-up all quantities have clear and intuitive physical meanings, such as vorticity, momentum and velocity field; the actual physics of the flow is present through the metric and volume form, which are essential but used sparingly, and no covariant derivatives are needed.
We then followed the general framework as far as we could using only the assumption of vortex slenderness to gain a complicated set of equations, which for convenience are set out in appendix \ref{appgensys}. These include equations for the motion of the centreline of the vortex, the curve $\Cc(t)$, in (\ref{eqcurvebasisdt1x}, \ref{eqS0defx}), equations for a straight, steady columnar vortex (\ref{eqgovlag4x}) giving the leading order local structure of the filament, and then first order equations (\ref{eqgovlag5x}, \ref{eqawkflucexp2x}) giving its evolution on longer space and time scales. We showed that this system includes both vortex ring dynamics with axial flow, and area waves on a vortex \citep{LuAs89,Le94,Ma02}.  

We will now assess the advantages and then disadvantages of our framework, and future directions. 
The systems of equations we have developed have the benefit that the vorticity equation is highly simplified: the axial vorticity is $\xib_3(r)$ is exactly conserved, while differential rotation creates azimuthal vorticity $\xib_2(r,z,t)$ and axial flow $\vb^3(r, z, t)$. 
Thus the Lagrangian conservation of vorticity is built in, while the geometric set-up concerns how this is embedded into three-dimensional space, whether in the early exact framework, or in the later slender approximation. 
There is then the potential for vorticity-preserving numerical methods to be developed out of either the exact or slender systems, in which time-stepping the mapping and metric are the key components. 
Both systems require the vorticity to have the appropriate topological structure in the $(r, \theta, z)$ coordinate system, depicted in say figure \ref{fig2}, but otherwise allow general streamline geometry within the vortex tube as defined by the metric coefficients $g_{ij}$. 
Thus, when modelling a slender vortex, there is no need to assume that it is nearly circular in cross section; the framework can absorb significant core distortions. 
Although a slender vortex moving in isolation will tend to have a nearly circular core, if vortices come close and start to interact, significant core distortions can occur, driven by the irrotational flow in which each core sits. 
For example, our framework has the potential to be applied to one lobe of a vortex dipole where two vortices are driven together in a hairpin configuration, including the axial flows that are driven by variation in core radius \citep{BuKe08, MoKi19a, MoKi19b, ChGi21, ChGi25}. 
The framework also allows the curve $\Cc(t)$ to have velocity $\Uv_{\cen}$ of order unity, with reference to our scalings, much greater than the $O(\eps)$ velocity from vortex self-induced motion, and again relevant if vortices start to interact. 
Finally the framework could perhaps be useful to analysts in establishing rigorous results on vortex configurations and their existence and stability, by allowing a focus on the properties of the geometrical embedding of the vorticity field in three-dimensional space.

Despite these advantages, there are numerous disadvantages. 
The main one is that the general framework is undoubtedly complicated even in the slender case. 
For example, there are challenges in tracking leading order and first order metric coefficients for cores that deviate significantly from being circular. 
Certainly in analytical modelling, the core structure would have to be reasonably straightforward to allow progress, for example allowing elliptical cores rather than of general shape. 
Allowing a model to follow the core dynamics in detail, as ours does through being vorticity preserving, has the further disadvantage that fluid dynamics within a vortex core can itself be complicated  and the slenderness assumption can rapidly be lost as axial flows develop \citep{MeHu94, ChGi21}. Some sort of smoothing or averaging may be needed to allow a focus on longer time scales, specific to an application. 
In any application there is likely also the need to bring in the Biot--Savart law, leading to a complex set of coupled PDEs and integrals. 
Another disadvantage is that the process of erosion of vortex dipoles, where vortex lines slip past a rear stagnation point and are lost to a thin layer in the flow \citep[e.g.,][]{Ri98, ChGiVa16}, appears to be common in many simulations \citep[e.g.,][]{BuKe08} but is difficult for us to incorporate. 
The erosion process represents either a loss of vorticity from a vortex --- but our modelling is vorticity-preserving --- or a secular distortion of the vorticity field which is painful to track in our coordinate system. In any case, action--angle coordinates become non-uniform in the presence of stagnation points and erosion. 

However, going back to advantages and forward to future directions, the framework is so general that it must contain numerous models for vortex filaments including those already in the literature, through applying further simplifications and approximations.  
As such it forms the basis on which to approach specific questions involving, for example, vortex waves and instabilities \citep{WiBlTs74, KoCh91, KoCh95}, interaction of hairpin vortices \citep{BuKe08, MoKi19a, MoKi19b, ChGi21, ChGi25}, and maybe vortex breakdown \citep{Le78}. 

We briefly outline how one might set about applying our work to study a particular vortex configuration. 
First, the general framework may well simplify initially because some quantities are not as large as could be allowed (for example this was the case for vortex rings, with $U = W = \tau = 0$, $V=O(\eps)$). 
We would then need a model for the local core structure of a slender vortex. 
Although the equations for this are written in a geometric form in (\ref{eqgovlag4x}), the best approach is probably to recognise that we are simply solving for what is locally a two-dimensional flow so need to satisfy $\nabla^2 \psi = F(\psi)$; we can then extract what metric information we need. 
We would likely choose the local core structure from some well-known family of solutions for two-dimensional flow given by some parameters $\{\lambda_i(z,t)\}$ that vary along the vortex. 
Going beyond approximately circular cores, we could for example consider elliptical vortices maintained by the external strain field obtained by application of the Biot--Savart law. 
Likewise a local axial flow could be chosen from a family of suitable candidates with further parameters in the set $\{\lambda_i(z,t)\}$. For example in the most basic modelling one might use the gauge freedom of setting $W_{\cen}$ to capture some measure of the mean axial flow, set the rest to zero, and so simplify the dynamics as done for area waves \citep{LuAs89,Le94, Ma02, ChGi21}. 
We would then go to the first order equations (\ref{eqgovlag5x}, \ref{eqawkflucexp2x}) and seek equations for the $\lambda_i(z,t)$, making use of further approximations and truncations. 
At some point such a model would become a system of PDEs in $(z,t)$, to be solved numerically and compared with simulation or experiment. 
We plan further investigations along these lines.


\vspace{5mm}

\noindent
{\bf Acknowledgements}

\noindent
I am ever grateful to Steve Childress for his insights into vortex dynamics, contour averaging, coordinate systems and asymptotic methods. 
I also thank Jacques Vanneste for numerous discussions on matters of geometry and fluid dynamics. 

\noindent
{\bf For the purpose of open access, the
author has applied a Creative Commons Attribution (CC BY) licence to any Author Accepted Manuscript
version arising from this submission.}

\noindent
{\bf Funding} 
    
\noindent 
Parts of this work were undertaken during the EPSRC grant EP/T023139/1, which I gratefully acknowledge.
 
\noindent
{\bf Declaration of interests}

\noindent
The author reports no conflicts of interest.

\noindent
{\bf Data access statement}

\noindent
No numerical codes or data sets were used in this study. 

\noindent
{\bf Author ORCID}

\noindent
Andrew D. Gilbert https://orcid.org/0000-0002-6940-1801;
 

{}


\appendix


\section{Outer flow field for vortex ring motion}\label{appouterflow}

\noindent
For completeness we will give the outer flow field needed for vortex ring motion, based on \cite{Sa92} and \cite{La94}, but using the notation of the present paper. The Stokes stream function $\Psi$ outside a thin vortex tube of radius $\ell$ centred at the origin of polar coordinates $(R, \Theta, Z)$ in the ambient space is given by:
\begin{align}
\Psi(R,Z) &  = \frac{\Gamma R\ell}{4\pi} \int_0^{2\pi} \frac{\cos \theta\, d\theta}{ ( Z^2 + R^2 + \ell^2 - 2 R \ell \cos \theta)^{1/2} } \\
 & = \frac{\Gamma R^{1/2} \ell^{1/2}}{2 \pi} \bigl[ ( 2k^{-1}-k) K(k) - 2k^{-1} E(k) \bigr] , 
\end{align}
at any point $(R, \Theta, Z)$, where we set (the prime here is just a label), 
\begin{align}
k^2 & = \frac{4 R \ell} {Z^2 + (R+  \ell )^2 } \, , \quad 
k'^2 = 1- k^2  = \frac{ Z^2 + (R - \ell)^2}{Z^2 + (R+\ell)^2}\,  ,
\end{align}
and $E$ and $K$ are elliptic integrals \citep[e.g.,][chapter 19]{OlLoBoCl10}. As the point $(R,Z)$ approaches the thin vortex we have $R\to \ell$ and $Z\to0$; in this limit $k'\to0$ and $\Psi$ may be approximated by 
\begin{align}
\Psi(R,Z)  & \simeq \frac{\Gamma R^{1/2} \ell^{1/2}}{2 \pi} \Bigl[ \, \log\, \frac{4}{k'} - 2 + O(k'^2 \log k' )\,\Bigr] .
\end{align}
This is in the ambient space coordinates. If we now move to the vortex coordinate system $(\rt, \thetat, \zt)$ we have 
\begin{equation}
R = \ell - \rt \cos \thetat, \quad Z = \rt \sin \thetat.
\end{equation}
Approximation for small $\rt$ yields
\begin{align}
\Psi & \simeq  \frac{\Gamma \ell}{2 \pi} \Bigl[ \, \log \frac{8 \ell}{\rt} - 2 -  \frac{\rt}{2\ell} \log \frac{8 \ell}{\rt} \cos \thetat +  \frac{\rt}{2 \ell} \cos \thetat \, \Bigr]  .
\end{align}
This gives the function of space $\Psi$ rewritten in terms of $(\rt,\thetat)$ coordinates, valid for small $\rt$, set out in (\ref{eqPsiouter1}). We pick up the discussion in \S\ref{ssecringvel}.

\details{
\begin{align}
k'^2 & = \frac{ \rt^2 }{\rt^2 -  4 \ell \rt \cos \thetat + 4 \ell^2 } , \quad 
k'  \simeq \frac{\rt}{2\ell} \bigl( 1 +  \frac{ \rt}{2\ell} \cos \thetat + \cdots\bigr) \\
R^{1/2} & \simeq \ell^{1/2} \bigl( 1 - \frac{\rt}{2\ell} \cos \thetat + \cdots\bigr) , \\
\Psi & \simeq  \frac{\Gamma \ell}{2 \pi} \Bigl[\,   1 - \frac{\rt}{2\ell} \cos \thetat + \cdots  \, \Bigr]
 \Bigl[ \, \log\, \frac{8\ell}{\rt} - 2 - \frac{\rt}{2\ell} \, \cos \thetat + \cdots \,\Bigr]
\end{align}
}


\section{Vortex ring motion for general vorticity profiles}\label{appgenprofile}

\noindent
In this appendix we revisit the calculation of the vortex ring velocity for general profiles. The homogeneous version of the key equation (\ref{eqringODE3}) has exact solutions 
\begin{equation}
f = r^{-1}, \quad  r^{-1} \int^r \frac{t}{\varsigma(t)^{2}} \, dt , 
\end{equation}
where it is convenient to set $\varsigma = r^2 \vb^2$ so that $\sigmab = r \varsigma_r / \varsigma$ in (\ref{eqdefsimgabrhob}) and we use $s$ and $t$, in this appendix only, as integration variables. 

The first of these solutions, $f = r^{-1}$, corresponds to an infinitesimal translation of the flow field. It is singular as $r\to0$, and this relates to the awkwardness of describing circular streamlines about a point displaced slightly from the origin in polar coordinates. Requiring that $f$ is finite as $r\to 0$ eliminates this solution and pins down the curve $\Cc$ as the geometric centre of the vorticity surfaces. Note that the interpretation of this solution and why it is there explains the ``fortuitous solution, by inspection, of the perturbed vorticity equations'', as \cite{WiBl71} describe part of the analysis for vortex core structure in \cite{WiBlZa71}, with analogies to our development but different in detail.

The Green's function given by 
\begin{equation}
G(r,s) = \frac{\varsigma(s)^2}{rs^2} \int_s^r  \frac{t}{\varsigma(t)^{2}}  \, dt  \quad \text{for}\quad r>s , 
\end{equation}
and $G(r, s)=0$ for $r<s$, provides a solution, 
\begin{equation}
f(r) =   \int_0^r G(r,s) \, \bigl(2\sigmab(s)  +1 + \rhob(s) \bigr)  \kappa s  \, ds ,
\label{eqfGreens}
\end{equation}
to (\ref{eqringODE3}) that is regular at the origin, $r=0$. The flow field outside the vortex gives constant $\varsigma(r) = \Gamma / 2\pi $, that is for $r>a$, where $\Gamma$ is the vortex circulation (equal to $2\pi \Omega a^2$ in our example (\ref{eqbasicstatezeta})). Thus we find that 
\begin{equation}
G(r,s)  = 
\begin{cases}
\displaystyle
\frac{1}{2r s^2}\, ( r^2 -s^2 )  &  \text{for}\quad r>s >a , \\[0.9em]
\displaystyle
\frac{\varsigma(s)^2}{rs^2}  \int_s^a  \frac{t}{\varsigma(t)^{2}}  \, dt 
+  \frac{4\pi^2}{\Gamma^2} \,\frac{\varsigma(s)^2}{2r s^2}  \,( r^2 - a^2 )  
 & \text{for}\quad r>a > s  .
 \end{cases}
\end{equation}
Substituting this into (\ref{eqfGreens}) gives (recalling that $\sigmab(r) =\rhob(r)= 0$ for $r>a$), 
\begin{align}
f(r)  &  =\tfrac{1}{2}   \kappa r \bigl[  \log ({r}/{a}) - \tfrac{1}{2}  + I  \bigr]   + O(r^{-1})  \tas r\to \infty, 
\end{align}
with 
\begin{align}
I & = \frac{ 4 \pi^2 }{\Gamma^2}  \int_0^a \varsigma(s)^2  \bigl[2\sigmab(s)  +1 + \rhob(s) \bigr] \,s^{-1} \, ds. 
\end{align}
Reusing $r$ as a dummy variable and the definitions of $\sigmab$ and  $\rhob$ in (\ref{eqdefsimgabrhob}), and $\varsigma$ above, allows this to be written as 
\begin{align}
I & =1 +  \frac{4 \pi^2}{\Gamma^2}  \int_0^a (r \vb^2)^2 \,  r \, dr 
-  \frac{8 \pi^2}{\Gamma^2}  \int_0^a   (\vb^3)^2  \, r\, dr . 
\end{align}
Putting this together with (\ref{eqVcenlimit}) gives the vortex velocity as 
\begin{equation}
V_\cen =  \frac{\Gamma \kappa}{4\pi} \, 
\Bigl(  \, \log \frac{8}{a\kappa} - \frac{1}{2} +  \frac{4 \pi^2}{\Gamma^2}  \int_0^a (r \vb^2)^2 \,  r \, dr 
-  \frac{8 \pi^2}{\Gamma^2}  \int_0^a   (\vb^3)^2  \, r\, dr \,\Bigr) , 
\label{eqVcengeneral}
\end{equation}
generalising (\ref{eqVcenspecific}), and involving a mean-square average of the azimuthal velocity $r\vb^2$ and axial velocity $\vb^3$ across the vortex cross section, in agreement with \cite{WiBlZa71} and \cite{Sa70, Sa92}. 

\details{Here we have used that the circulation is $\Gamma = 2\pi \varsigma(a)$; the second integral arises after integration by parts using that $\vb^3(a)=0$.}


\section{Complete general system}\label{appgensys}

\noindent
Here we set out the full final governing system obtained, removing extraneous material, in one place. We drop the subscript `$\oo$' from fields such as $\vb^2_{\oo}$ to make the point that we have sought to obtain all equations that do not involve any first order fields. 

We have the centre line equations (\ref{eqcurvebasisdt1}), 
\begin{subequations}\label{eqcurvebasisdt1x}
\begin{align}
\partial_{t} J |_{z} & = J(W_{\cen\arcl} - \kappa U_\cen) ,  \\
\partial_{t} \kappa |_{z} & = (U_{\cen\arcl} - \tau V_\cen)_\arcl  - \tau(V_{\cen\arcl} + \tau U_\cen)    + \kappa_\arcl W_\cen + U_\cen \kappa^2 , \\
\partial_{t} \tau  |_{z} & = S_{\cen\arcl}  + \kappa (V_{\cen\arcl} + \tau U_\cen) - \tau(W_{\cen\arcl} - \kappa U_{\cen}), 
\end{align}
\end{subequations}
with (\ref{eqS0def}), 
\begin{equation}
S_\cen  = \kappa^{-1} (V_{\cen\arcl} + \tau U_\cen)_\arcl + \kappa^{-1}\tau ( U_{\cen\arcl} - \tau V_\cen ) + \tau W_\cen. 
\label{eqS0defx} 
\end{equation}
The equations that govern columnar vortex dynamics at leading order, independent of curvature, are taken from (\ref{eqgovlag4}), 
\begin{subequations}\label{eqgovlag4x}
 \begin{align}
\xib_{3}(r) & = \partial_r \nub_{2}(r), \\
\xib_{2} & =   - \partial_r \nub_{3}, 
\\
\nub_{2} & = \gb_{22\oo} \vb^2  , 
\\
\nub_{3} - \Upsilonb_{3\oo} & =  \gb_{33\oo} \vb^3 , 
\\
 0
& = (\partial_r g'_{22\oo} - \partial_\theta g'_{12\oo}) \vb^2 + g'_{22\oo} \partial_r \vb^2 ,
\\
 \alpha'_{\oo}  & = 0 . 
\end{align}
\end{subequations}
The fields evolve on the slow $(z,t)$ scales according to the first order equations from (\ref{eqgovlag5}), 
\begin{subequations}\label{eqgovlag5x}
 \begin{align}
 \partial_{t}  \xib_{2} & + \partial_{z} (\vb^3 \xib_{2}  )  =  \xib_{3} \,  \partial_{z} \vb^2 , 
 \\
- \Xi'_{1\ii}
& = (\partial_\theta g'_{32\ii} - \partial_z g'_{22\oo}) \vb^{2} 
+(\partial_\theta  g'_{33\ii}) \vb^3
- g'_{22\oo} \partial_z \vb^2 , 
\\
- \Xi'_{2\ii} 
& = \partial_z (g'_{12\oo} \vb^2 ) 
- \partial_r ( g'_{32\ii} \vb^2 +  g'_{33\ii} \vb^3  ) , \\
\partial_t \mub_{\oo}   & +  \partial_z (\mub_{\oo} \vb^3)  = 0 , \\
\mu'_{\ii} & = 0 , 
\end{align}
\end{subequations}
together with (\ref{eqawkflucexp2}), 
 \begin{align}
   \Bigl (\, \frac{1}{g'_{22\oo}} \, 
  \bigl[  \Xi'_{3\ii}+ (\partial_r g'_{22\ii} - \partial_\theta g'_{12\ii}) \vb^2  
+(\partial_r g'_{23\ii} - \partial_\theta g'_{13\ii}) \vb^3  
+ g'_{22\ii} \partial_r \vb^2  + g'_{23\ii} \partial_r \vb^3 
\bigr] \,\Bigr)' 
& = 0 
\label{eqawkflucexp2x} 
\end{align}
(or, if $g'_{22\oo}=0$ then we revert to consideration of (\ref{eqgovlag5}d)). 
Equations (\ref{eqgovlag5x}b, c) are not independent since the relationship (\ref{eqdXiprime1}) holds,
\begin{equation}
\partial_r \Xi'_{1\ii} + \partial_\theta \Xi'_{2\ii} = 0 .
\label{eqdXiprime1x}
\end{equation}

These governing systems involve the component $\Upsilonb_{3\oo}$ of the background flow momentum, 
from (\ref{eqUpsilonbasic1}),
\begin{align}
\Upsilonb_{3\oo}  &  = JW_\cen , 
\label{eqUpsilonbasic1x}
\end{align}
and the background vorticity components $\Xi'_{1\ii}$, $\Xi'_{2\ii}$, $\Xi'_{3\ii}$, which can be read off as the fluctuating components of (\ref{eqZetabasic2}), for $\Xiv = \Xiv_{\ii} + \cdots$, namely 
\begin{subequations}\label{eqZetabasic2x}
\begin{align}
\Xi_1  &   
= - 2J   (U_{\cen s} - \tau V_\cen)   \xt_\theta -   2 J (V_{\cen s}+ \tau U_\cen) \yt_\theta   - 2 J \kappa W_\cen \xt_\theta + \cdots , 
  \\
\Xi_2  &  
=   2J   (U_{\cen s} - \tau V_\cen)   \xt_r + 2 J (V_{\cen s}+ \tau U_\cen) \yt_r    + 2 J  \kappa W_\cen \xt_r + \cdots, \\
\Xi_3 &  
=  2 \alpha S_\cen   +  \xt_\theta \xt_{rt} - \xt_r \xt_{\theta t}  + \yt _\theta \yt_{r t}-  \yt _r \yt_{\theta t}    + \cdots.
\end{align}
\end{subequations}
Metric information can be obtained from (\ref{eqmetricslender}), namely
\begin{equation}\label{eqmetricslenderx}
(g_{ij}) =  
\begin{pmatrix}
\xt_r^2 + \yt_r^2 &\; \xt_r \xt_\theta + \yt_r \yt_\theta \; & \xt_r \xt_z + \yt_r \yt_z  + J \tau ( \xt \yt_r - \yt \xt_r) + J^2 q_r \\
\otimes & \xt_\theta^2 + \yt_\theta^2 &   \xt_\theta \xt_z + \yt_\theta \yt_z + J \tau ( \xt \yt_\theta - \yt \xt_\theta) + J^2 q_\theta \\
\otimes & \otimes& J^2h^2
 \end{pmatrix}
 +\cdots , 
 \end{equation}
with (\ref{eqhdef}), 
\begin{equation}
h = 1 - \kappa \xt  .
\label{eqhdefx}
\end{equation}
The quantities $\alpha_{\oo}$ and $\alpha_{\ii}$ can be found from expanding (\ref{eqalphaslender1}), which is
\begin{equation}
\alpha^2 = g_{11}g_{22} - g_{12} g_{12}, 
\label{eqalphaslender1x}
\end{equation}
while $\mu_{\ii}$ can be obtained from (\ref{eqmuslender}) 
\begin{equation}
\mu =     J  h \alpha + \cdots, 
\label{eqmuslenderx}
\end{equation}
specifically $\mu_{\ii} =     J  ( \alpha_{\ii} -  \kappa \alphab_{\oo} \xt_{\oo})$.
Solutions need to be matched to an external irrotational flow, presumably provided by application of the Biot--Savart law. We finally note that there remains the gauge freedom in (\ref{eqgauge1}, \ref{eqguageagain}), particularly of selecting $W_{\cen}$, for example to create simplified models of axial flow. 
%

 



\end{document}

%% file: latex-macros.tex
 \usepackage{natbib}

\usepackage{graphicx,color}

\usepackage{amssymb}
\usepackage{amsmath}
\usepackage{amsfonts}

\usepackage{multirow}
\usepackage{subeqnarray}

\numberwithin{equation}{section}

\newcommand{\rt}{\tilde{r}}

\newcommand{\ttt}{\tilde{t}}

\newcommand{\xt}{\tilde{x}}
\newcommand{\yt}{\tilde{y}}
\newcommand{\zt}{\tilde{z}}

\newcommand{\thetat}{\tilde{\theta}}

\newcommand{\bv}{\boldsymbol{b}}

\newcommand{\gv}{\boldsymbol{g}}
\newcommand{\iv}{\boldsymbol{i}}
\newcommand{\jv}{\boldsymbol{j}}
\newcommand{\kv}{\boldsymbol{k}}

\newcommand{\nv}{\boldsymbol{n}}
\newcommand{\rv}{\boldsymbol{r}}

\newcommand{\tv}{\boldsymbol{t}}
\newcommand{\uv}{\boldsymbol{u}}
\newcommand{\vv}{\boldsymbol{v}}

\newcommand{\Qv}{\boldsymbol{Q}}

\newcommand{\Uv}{\boldsymbol{U}}

\newcommand{\alphav}{\boldsymbol{\alpha}}

\newcommand{\psiv}{\boldsymbol{\psi}}
\newcommand{\omegav}{\boldsymbol{\omega}}
\newcommand{\muv}{\boldsymbol{\mu}}
\newcommand{\nuv}{\boldsymbol{\nu}}

\newcommand{\piv}{\boldsymbol{\pi}}
\newcommand{\varpiv}{\boldsymbol{\varpi}}

\newcommand{\xiv}{\boldsymbol{\xi}}

\newcommand{\Psiv}{\boldsymbol{\Psi}}
\newcommand{\Xiv}{\boldsymbol{\Xi}}
\newcommand{\Upsilonv}{\boldsymbol{\Upsilon}}

\newcommand{\tth}{\hat{t}}

\newcommand{\zh}{\hat{z}}

\newcommand{\Jh}{\hat{J}}
\newcommand{\Wh}{\hat{W}}

\newcommand{\thetah}{\hat{\theta}}

\newcommand{\gb}{\bar{g}}

\newcommand{\ub}{\bar{u}}
\newcommand{\vb}{\bar{v}}

\newcommand{\alphab}{\bar{\alpha}}

\newcommand{\nub}{\bar{\nu}}
\newcommand{\sigmab}{\bar{\sigma}}
\newcommand{\rhob}{\bar{\rho}}
\newcommand{\psib}{\bar{\psi}}
\newcommand{\mub}{\bar{\mu}}
\newcommand{\xib}{\bar{\xi}}

\newcommand{\Upsilonb}{\bar{\Upsilon}}
\newcommand{\Xib}{\bar{\Xi}}

\newcommand{\uvb}{\bar{\uv}}
\newcommand{\vvb}{\bar{\vv}}
\newcommand{\xivb}{\bar{\xiv}}

\usepackage{eucal}  

\newcommand{\Cc}{{\mathcal{C}}}

\newcommand{\Sc}{{\mathcal{S}}}
\newcommand{\Tc}{{\mathcal{T}}}

\newcommand{\Lc}{\mathcal{L}}

\newcommand{\Rbb}{\mathbb{R}}

\DeclareMathOperator{\divv}{div}

\newcommand{\eps}{\varepsilon}

\newcommand{\lie}{\mathcal{L}}
\newcommand{\ip}{\lrcorner}

\ifdefined\half
\else
\newcommand{\half}{\tfrac{1}{2}}
\fi

\ifdefined\flagforexam
\else

\fi

\newcommand{\tas}{\quad\text{as}\quad}

\newcommand{\tor}{\quad\text{or}\quad}

\usepackage{listings}
\usepackage{multirow}
\usepackage{xcolor} 

\definecolor{codegreen}{rgb}{0,0.6,0}
\definecolor{codegray}{rgb}{0.5,0.5,0.5}
\definecolor{codepurple}{rgb}{0.58,0,0.82}
\definecolor{backcolour}{rgb}{0.98,0.98,0.95}

\lstdefinestyle{mystyle}{
    backgroundcolor=\color{backcolour},   
    commentstyle=\color{codegreen},
    keywordstyle=\color{magenta},
    numberstyle=\tiny\color{codegray},
    stringstyle=\color{codepurple},
    basicstyle=\ttfamily\footnotesize\color{black},
    breakatwhitespace=false,         
    breaklines=true,                 
    captionpos=b,                    
    keepspaces=true,                 
    numbers=none,                    
    numbersep=5pt,                  
    showspaces=false,                
    showstringspaces=false,
    showtabs=false,                  
    tabsize=2
}
